\documentclass[aps,english,reprint,twocolumn,prd]{revtex4-1} 

\usepackage[utf8]{inputenc}
\usepackage[T1]{fontenc}
\usepackage[colorlinks]{hyperref}

\usepackage{amsmath,amssymb,amsfonts}
\usepackage{bbm}
\usepackage{babel}
\usepackage{graphicx}
\usepackage{color}
\usepackage{xcolor}
\usepackage{tensor}
\usepackage{slashed}
\usepackage[normalem]{ulem}

\newcounter{pointcounter}
\newcommand{\point}[1][\empty]{
  \ifthenelse
    {\equal{#1}{\empty}}
    {\ensuremath{\, \phantom{\cdot} \,}}
    {\setcounter{pointcounter}{1} \forloop{pointcounter}{0}{\value{pointcounter} < #1}{\ensuremath{\, \phantom{\cdot} \,}}}
}

\newcommand{\mcC}{\ensuremath{\mathcal{C}}}

\newcommand{\mcO}{\ensuremath{\mathcal{O}}}

\makeatother

\newcommand{\UV}{{\small UV}}

\newcommand{\RG}{{\small RG}}

\newcommand{\Rb}{{\bar{R}}}
\newcommand{\Sb}{{\bar{S}}}
\newcommand{\Cb}{{\bar{C}}}
\newcommand{\Db}{{\bar{D}}}
\newcommand{\gb}{{\bar{g}}}
\newcommand{\Deltab}{{\bar{\Delta}}}

\newcommand{\eg}{{\textit{e.g.}}}
\newcommand{\ie}{{\textit{i.e.}}}

\graphicspath{{./figures/}}

\begin{document}
\allowdisplaybreaks

\title{Correlation functions on a curved background}
\author{Benjamin Knorr}
\email{benjamin.knorr@uni-jena.de}
\affiliation{Theoretisch-Physikalisches Institut, Friedrich-Schiller-Universit\"at Jena, 
Max-Wien-Platz 1, 07743 Jena, Germany}

\author{Stefan Lippoldt}
\email{s.lippoldt@thphys.uni-heidelberg.de}
\affiliation{Institut f\"ur Theoretische Physik, Universit\"at Heidelberg, Philosophenweg 16, 69120 Heidelberg, Germany}

\begin{abstract}
We investigate gravitational correlation functions in a curved background with the help of
nonperturbative renormalization group methods.
Beta functions for eleven couplings are derived, two of which correspond to running gauge parameters.
A unique ultraviolet fixed point is found, suitable for a UV completion in the sense of
Asymptotic Safety.
To arrive at a well-behaved flow in a curved background, the regularization must be chosen carefully.
We provide two admissible choices to solve this issue in the present approximation.
We further demonstrate by an explicit calculation
that the Landau limit is a fixed point also for quantum gravity, and additionally show that in this limit,
the gauge parameter $\beta$ does not flow.
\end{abstract}

\maketitle

\section{Introduction}\label{sec:intro}

For several decades, Einstein's general relativity has successfully precluded any attempt of quantization.
At the heart of the problem lies the negative mass dimension of 
Newton's constant, which implies the failure of standard perturbation theory.
An alternative route was suggested by Weinberg \cite{Weinberg:1980gg}.
He proposed that gravity might be interacting in the far ultraviolet,
but controlled by a fixed point of its renormalization group (\RG{}) flow.
Such an interacting, or nontrivial fixed point is termed asymptotically safe,
in contrast to an asymptotically free fixed point, where the couplings vanish.

With the advent of modern functional \RG{} equations \cite{Wetterich:1992yh, Ellwanger1994, Morris:1993qb},
the Asymptotic Safety scenario received growing attention.
Starting with the seminal paper \cite{Reuter:1996cp}, in which the nonperturbative beta functions for
Newton's constant and the cosmological constant were derived for the first time,
approximations were successively improved.
This includes aspects of the Einstein-Hilbert approximation
\cite{Gies:2015tca, Souma:1999at, Lauscher:2001ya, Lauscher:2001rz, Reuter:2001ag, Litim:2003vp, Lauscher:2005qz,
Reuter:2005bb, Niedermaier:2006wt, Groh:2010ta, Benedetti:2010nr, Manrique:2011jc, Reuter:2012id,
Harst:2012ni, Litim:2012vz, Nink:2012kr, Rechenberger:2012dt, Nink:2014yya, Falls:2015qga, Falls:2015cta},
higher derivative terms \cite{Lauscher:2002sq, Codello:2006in, Benedetti:2009rx, Groh:2011vn, Rechenberger:2012pm, Ohta:2013uca, Ohta:2015zwa, Hamada:2017rvn, 
Christiansen:2016sjn},
$f(R)$ gravity \cite{Machado:2007ea, Codello:2007bd, Bonanno:2010bt, Dietz:2012ic, Demmel:2012ub, Falls:2013bv, Dietz:2013sba, Falls:2014tra, Demmel:2014hla,
Demmel:2014sga, Eichhorn:2015bna, Demmel:2015oqa, Ohta:2015efa, Ohta:2015fcu, Falls:2016wsa, Falls:2016msz, Morris:2016spn, Gonzalez-Martin:2017gza},
the resolution of the fate of the perturbative counterterm \cite{Gies:2016con}, the quantization of {\small{ADM}} variables \cite{Biemans:2016rvp, 
Biemans:2017zca, Houthoff:2017oam}, the inclusion of torsion and nonmetricity \cite{Pagani:2015ema} and progress on unitarity \cite{Nink:2015lmq}.
Recently, there is growing interest in the study of gravity-matter systems
\cite{Percacci:2002ie, Percacci:2003jz, Daum:2009dn, Vacca:2010mj, Folkerts:2011jz, Eichhorn:2011pc,
Harst:2011zx, Eichhorn:2012va, Dobrich:2012nv, Dona:2012am, Dona:2013qba, Percacci:2015wwa, Borchardt:2015rxa,
Meibohm:2015twa, Dona:2015tnf, Labus:2015ska, Meibohm:2016mkp, Eichhorn:2016esv, Eichhorn:2016vvy, Christiansen:2017gtg, Christiansen:2017qca, 
Eichhorn:2017eht, Biemans:2017zca}.
All studies come to the conclusion that there is indeed a suitable fixed point
which facilitates an ultraviolet (\UV{}) completion of gravity.
Phenomenologically, black holes \cite{Bonanno:2000ep, Falls:2010he, Falls:2012nd, Becker:2012js, Koch:2013owa, Koch:2014cqa},
cosmological aspects \cite{Bonanno:2001xi, Reuter:2005kb, Bonanno:2007wg, Hindmarsh:2011hx, Henz:2013oxa, Saltas:2015vsc, Bonanno:2015fga, Henz:2016aoh, 
Bonanno:2016dyv, Wetterich:2017ixo, Bonanno:2017pkg},
the Unruh effect \cite{Alkofer:2016utc}, the C-function \cite{Becker:2014pea}
and the dispersion of different modes \cite{Biemans:2016rvp} have been investigated.
Assuming that there is an asymptotically safe fixed point, a precise prediction of the Higgs mass
was made in \cite{Shaposhnikov:2009pv} before its measurement at the {\small{LHC}} \cite{Chatrchyan:2012xdj, Aad:2012tfa}.

For technical reasons, the background field method is indispensable in these calculations.
In this, the metric is split into a background and a (not necessarily small) fluctuation field.
It was soon realized that the disentanglement of these two quantities is central to obtain reliable results.
In \cite{Litim:2002ce, Folkerts:2011jz} it was shown that an improper treatment of this difference can alter universal one-loop
beta functions, and even destroy asymptotic freedom in Yang-Mills theory. Similarly, the well-known Wilson-Fisher fixed point can disappear \cite{Bridle:2013sra}.
To solve this problem, one has to deal either with two fields \cite{Manrique:2010mq, Manrique:2009uh, Manrique:2010am,
Christiansen:2012rx, Christiansen:2014raa, Becker:2014qya, Becker:2014jua, Christiansen:2015rva, Meibohm:2015twa,
Dona:2015tnf, Meibohm:2016mkp, Denz:2016qks, Henz:2016aoh, Christiansen:2016sjn},
or solve the corresponding split Ward identities \cite{Pawlowski:2003sk, Pawlowski:2005xe, Donkin:2012ud, Bridle:2013sra, Dietz:2015owa, Labus:2016lkh,
Morris:2016nda, Morris:2016spn, Percacci:2016arh, Ohta:2017dsq}.
Closely related are geometric quantization schemes
\cite{Donkin:2012ud, Demmel:2014hla, Wetterich:2016ewc, Wetterich:2016qee}.

In the study of fluctuation correlation functions, so far the analysis was restricted to a flat background.
This bears technical advantages, \eg{}, the full momentum dependence can be resolved \cite{Christiansen:2014raa}.
However, it is the functional dependence of the effective action on the background field which is necessary for the calculation of observables 
\cite{Branchina:2003ek}.
Hence, the introduction of a generic background is unavoidable and the study of correlation functions including the background curvature is important.
As a first step, we resolve the curvature dependence of the graviton propagator
to linear order in the curvature within a derivative expansion.
This is a further step in the systematic exploration of correlation functions in quantum gravity,
enabling us to assess in what way quantum effects in the \UV{} shift the dependence
of the propagator on the curvature compared to the classical expectation from the Einstein-Hilbert action.

We show within our truncation that the Landau gauge is a fixed point of the \RG{} flow of both gauge parameters.
This may seem to be a trivial statement, but it turns out that a careful choice of the regulator is necessary to obtain a finite right-hand side of the flow 
equation \cite{Gies:2015tca}. It is in general difficult to find a valid choice on a curved background geometry, since functions of the Laplacean don't commute 
with covariant derivatives. We provide two different admissible regulators to linear order in the background curvature.
One of the choices does not involve the gauge parameter $\alpha$, and thus is technically superior as it allows to take the Landau limit at 
the level of the propagator. Furthermore, it is straightforward to extend this regulator to higher order truncations. The other regulator is the curved version 
of the regulator as employed in \cite{Christiansen:2012rx, Christiansen:2014raa, Christiansen:2015rva, Denz:2016qks, Christiansen:2016sjn}. For fluctuation 
flows on flat background 
in Landau limit, both regulators give the same result.
To judge the quality of our approximation, we study the beta functions for general gauge parameters,
similar to our earlier work within the background field approximation \cite{Gies:2015tca}.

In Landau gauge, we find one fixed point which still depends somewhat on the remaining gauge parameter $\beta$. For some choices, this fixed point is 
\UV{}-repulsive, where the leading critical exponents have large imaginary parts, which we take as a hint that the inclusion of higher order correlation 
functions is necessary to ultimately fix the critical quantities.

This work is structured as follows: in section \ref{sec:corrfuncs}, we give the basic notions of our \RG{} setup,
together with the employed truncation in subsection \ref{subsec:truncation},
the regularization in subsection \ref{subsec:reg} and the projection scheme,
together with a discussion of the Landau gauge in subsection \ref{subsec:proj}.
We go on with the discussion of the results in section \ref{sec:results}, where we first consider the 0th order curvature couplings
in subsection \ref{subsec:flat}, then the 1st order curvature couplings in subsection \ref{subsec:curved}.
We end with a conclusion in section \ref{sec:summary}.
The appendices collect some technical information.
In appendix \ref{app:basis}, we give a basis for a set of correlation functions, whereas in appendix 
\ref{app:prop}, we
give some helpful relations concerning the propagator functions
for a symmetric spin 2 field. In appendix \ref{app:regcomp}, we give explicit 
fixed point values for the two different regulators.

\section{Nonperturbative correlation functions in quantum gravity}\label{sec:corrfuncs}

A theory is completely fixed if a complete set of correlation functions is given, as any observable can be constructed from these basic building blocks. These 
correlation functions are generated by the effective action at vanishing fluctuation field. To study the effective action nonperturbatively, we use the 
formulation of the functional \RG{} by Wetterich \cite{Wetterich:1992yh}. For this, a fiducial scale $k$ is introduced, and momentum shells are integrated out 
at this scale successively in a Wilsonian sense.
The $k$-dependent, so-called effective average action, $\Gamma$,
fulfills the \RG{} equation
\begin{equation}\label{eq:flowequation}
 \dot \Gamma \equiv k \partial_k \Gamma = \frac{1}{2} \text{STr} \left[ \left( \Gamma^{(2)} + \mathfrak R \right)^{-1} \, k \partial_k \mathfrak{R} \right] \, .
\end{equation}
In this equation, STr indicates a supertrace, which includes summation over discrete and integration over
continuous indices as well as a minus sign for Gra\ss{}mann-valued fields,
and $\mathfrak R$ is a regulator, which effectively behaves like a momentum-dependent mass term. Reviews of the functional \RG{} in gravity can be found in 
\cite{Reuter:1996ub,Pawlowski:2005xe,Niedermaier:2006wt,Percacci:2007sz,Reuter:2012id,Nagy:2012ef}.

\subsection{Truncation}\label{subsec:truncation}

In the following we present our truncation scheme to solve \eqref{eq:flowequation}. As a starting point, we take the Einstein-Hilbert action,
\begin{equation}
 S_\text{cl} = \frac{1}{16\pi G_N} \int \sqrt{g} \left( - R + 2 \Lambda \right) \, ,
\end{equation}
where $G_N$ is the classical Newton's constant, $\Lambda$ the cosmological constant and $R$ the Ricci scalar of the metric $g$. We implement the background 
field method by a linear split,
\begin{equation}
 g_{\mu\nu} = \bar g_{\mu\nu} + h_{\mu\nu} \, .
\end{equation}
Other choices are also possible, see \eg{} \cite{Nink:2014yya, Demmel:2015zfa, Percacci:2015wwa, 
Gies:2015tca, Labus:2015ska, Ohta:2015zwa, Ohta:2015efa, Falls:2015qga, Ohta:2015fcu, Dona:2015tnf, Ohta:2016npm, Falls:2016msz, Percacci:2016arh}. Our goal is 
to resolve the flow of the two-point-correlator $\Gamma^{(2)}$, including up to two derivatives or one curvature, and parts of the three-graviton vertex 
$\Gamma^{(3)}$ similar to \cite{Christiansen:2015rva}.
Let us start by parameterizing the inverse propagator.
For this, we amend the quadratic part of the classical action with a gauge fixing and further couplings, which allows us to go beyond the background field 
approximation. The constant part can be spanned by two gaps, $\Lambda_\text{TL}$ and $\Lambda_\text{Tr}$, corresponding to the traceless and the trace sector 
of the fluctuation, respectively. To linear order in the background curvature, five independent tensor structures appear, and we supply each of them with a 
unique coupling $\mathcal R$. These couplings are introduced in such a way that their classical value is zero. Finally, we introduce a uniform wave function 
renormalization $Z_h$.

The gauge fixing is given by
\begin{equation}
 F_\mu = \left( \delta^{(\alpha}_\mu \bar D^{\beta)} - \frac{1+\beta}{4} \bar g^{\alpha\beta} \bar D_\mu \right) g_{\alpha\beta} \, .
\end{equation}
Here, $\bar D$ denotes the covariant derivative with respect to the background metric $\bar g$, whereas $D$ in the following corresponds to the covariant 
derivative with respect to the full metric $g$. The gauge fixing is implemented in a standard way by the Faddeev-Popov construction,
\begin{equation}
 \Gamma_\text{gf} = \frac{1}{16\pi G_N \alpha} \int \sqrt{\bar g} \, \bar g^{\mu\nu} F_\mu F_\nu \, ,
\end{equation}
which gives also rise to the ghost action
\begin{equation}
 \Gamma_\text{gh} = - \int \sqrt{\bar g} \, \bar c_\mu \left[ 2 \bar g^{\mu(\alpha} \bar D^{\beta)} - \frac{1+\beta}{2} \bar g^{\alpha\beta} \bar D^\mu \right] 
D_\alpha c_\beta \, .
\end{equation}
Our ansatz for the quadratic part of the effective action thus amounts to
\begin{widetext}
\begin{equation}\label{eq:action_quad}
\begin{aligned}
 \Gamma_\text{quad} &= \frac{Z_h}{64\pi} \int \sqrt{\bar g} \, h_{\mu\nu} \Big[ {\Pi_\text{TL}^{\mu\nu}}_{\rho\sigma} \left( \Deltab{} + \tfrac{2}{3} \left(  1 
+ 3 \mathcal R_\text{RTL} \right) \Rb{} - 2 \Lambda_\text{TL} \right) + 2 \left( \mathcal R_\text{C} - 1 \right) \tensor{\Cb{}}{^{\mu}_{\rho}^{\nu}_{\sigma}}
+2 \mathcal R_\text{STL} {\Pi_\text{TL}^{\mu\nu}}_{\alpha\beta} \Sb{}^\alpha_\gamma {\Pi_\text{TL}^{\beta\gamma}}_{\rho\sigma} \\
& \qquad\qquad\qquad\qquad + \frac{2(\alpha-1)}{\alpha} {\Pi_\text{TL}^{\mu\nu}}_{\alpha\beta} \bar D^\alpha \bar D_\gamma 
{\Pi_\text{TL}^{\beta\gamma}}_{\rho\sigma} + \frac{\beta-\alpha}{\alpha} \bar g^{\mu\nu} \bar D_\alpha \bar D_\beta 
{\Pi_\text{TL}^{\alpha\beta}}_{\rho\sigma} + 4 \mathcal R_\text{STr} \bar g^{\mu\nu} \Sb{}_{\alpha\beta} {\Pi_\text{TL}^{\alpha\beta}}_{\rho\sigma} \\
& \qquad\qquad\qquad\qquad + \frac{1}{2\alpha} {\Pi_\text{Tr}^{\mu\nu}}_{\rho\sigma} \left( \left( \beta^2 - 3 \alpha \right) \Deltab{} + 4 \alpha \left( 
\Lambda_\text{Tr} + \mathcal R_\text{Tr} \Rb{} \right) \right) \Big] h^{\rho\sigma} \, ,
\end{aligned}
\end{equation}
\end{widetext}
where we introduced the background Laplacean $\Deltab{} = -\bar D^\alpha \bar D_\alpha$.
Everything is spanned in a traceless decomposition.
In particular, we use the Weyl tensor $C$ and the trace-free Ricci tensor $S$ to rewrite the Riemann tensor,
\begin{equation}
 R_{\mu\nu\rho\sigma} = C_{\mu\nu\rho\sigma} + g_{\mu[\rho} S_{\sigma]\nu} + g_{\nu[\sigma} S_{\rho]\mu} + \frac{1}{6} R g_{\mu[\rho} g_{\sigma]\nu} \, ,
\end{equation}
and the Ricci tensor,
\begin{equation}
 R_{\mu\nu} = S_{\mu\nu} + \frac{1}{4} R g_{\mu\nu} \, .
\end{equation}
The projectors $\Pi_\text{TL}$ and $\Pi_\text{Tr}$ are defined as
\begin{equation}
\begin{aligned}
 {{\Pi_\text{TL}}^{\mu\nu}}_{\rho\sigma} &= {\mathbbm 1^{\mu\nu}}_{\rho\sigma} - {{\Pi_\text{Tr}}^{\mu\nu}}_{\rho\sigma} \, , \\
 {{\Pi_\text{Tr}}^{\mu\nu}}_{\rho\sigma} &= \frac{1}{4} \bar g^{\mu\nu} \bar g_{\rho\sigma} \, , \\
 {\mathbbm 1^{\mu\nu}}_{\rho\sigma} &= \frac{1}{2} \left( \delta^\mu_\rho 
\delta^\nu_\sigma + \delta^\mu_\sigma \delta^\nu_\rho \right) \, .
\end{aligned}
\end{equation}
It is useful to introduce a redefined gap parameter instead of $\Lambda_\text{Tr}$, which accounts for the fact that the two (off-shell) scalar degrees of 
freedom of the graviton mix, depending on the gauge fixing. Defining
\begin{equation}\label{eq:lambdatilde}
 \tilde\Lambda = \frac{6 \Lambda_\text{Tr} - 2 \beta^2 \Lambda_\text{TL}}{\left( \beta - 3 \right)^2} \, ,
\end{equation}
all propagators in the Landau limit have the denominator structure $\Deltab - 2 \Lambda_2$, with $\Lambda_2$ either being 
$\Lambda_\text{TL}$ or $\tilde\Lambda$. In the results below one can see that this parameterization is reasonable since fixed point values of these quantities 
change only mildly under the variation of the gauge parameter $\beta$, whereas $\Lambda_\text{Tr}$ shows a strong gauge dependence.

For the higher order correlation functions, we use the classical tensor structure as a model, in complete analogy to \cite{Christiansen:2015rva}.
In particular, this entails
\begin{align}
\begin{aligned} 
 \Gamma_\text{cub} &= \left. Z_h^{3/2} \, G_{3}^{1/2} \, G_N \, S_\text{cl, cub} \right|_{\Lambda\to\Lambda_3} \, ,
 \\
 \Gamma_\text{quart} &= \left. Z_h^2 \, G_4 \, G_N \, S_\text{cl, quart} \right|_{\Lambda\to\Lambda_4} \, ,
 \\
 \Gamma_\text{quint} &= \left. Z_h^{5/2} \, G_5^{3/2} \, G_N \, S_\text{cl, quint} \right|_{\Lambda\to\Lambda_5} \, ,
\end{aligned}
\end{align}
for the terms cubic, quartic and quintic in the fluctuation field $h$, respectively. 
The $G_n$ uniformly parameterize the interaction strength of the $n$-graviton vertex, while the $\Lambda_n$ characterize their constant parts.
In order to close the flow equation, we identify the couplings of the four- and five-point correlators with those of the three-point correlator, \ie{}
\begin{equation}
 G_5 = G_4 = G_3 \, , \qquad \qquad \Lambda_5 = \Lambda_4 = \Lambda_3 \, .
\end{equation}
A complete basis of all correlation functions up to third order in the fluctuation, second order in derivatives and first order in curvature can be found in 
appendix \ref{app:basis}. The strength of the ghost-graviton-vertex is also approximated by the same coupling $G_3$.

\subsection{Regularization}\label{subsec:reg}

On a general background, it is nontrivial to find a regulator such that also the curvature part of the 
flow stays finite, since there is a subtle interplay between the placing of the shape functions and the operators to be regularized. As a validity 
criterion for a given regulator, we demand that the Landau limit of the regularized propagator exists, since this is also the case for the unregularized 
propagator, and seems to be a natural requirement.

In appendix \ref{app:prop} we 
derive the general propagator of a symmetric spin 2 field on a flat background. Taking this as a starting point, we observe that for small momenta (\ie{} small 
eigenvalues of the Laplacean) only the prefactors of $\Pi_\text{TL}$ and $\Pi_\text{Tr}$ appear in denominators, and at least they need regularization. This 
inspires the following choice for the regulator:
\begin{equation}\label{eq:reg_minimal}
\begin{aligned}
 \Delta S_h \! &= \! \frac{Z_h}{64\pi} \int \!\!\! \sqrt{\bar g} \, h_{\mu\nu} \Big[ {\Pi_\text{TL}^{\mu\nu}}_{\rho\sigma} \! - \! \tfrac{5+\beta(\beta-2)}{2} 
{\Pi_\text{Tr}^{\mu\nu}}_{\rho\sigma} \Big] \mathfrak R(\Deltab{}) h^{\rho\sigma} .
\end{aligned}
\end{equation}
The argument is similar for the ghost field, which is a vector: only the prefactor of the identity appears in denominators, thus we can regularize
\begin{equation}
 \Delta S_c = \int \sqrt{\bar g} \, \bar c_\mu \mathfrak R(\Deltab{}) c^\mu \, .
\end{equation}
This regulator is well-behaved in the Landau limit by construction, and the prefactors are arranged such that in this limit, the denominators have the 
canonical form $\Deltab + \mathfrak R(\Deltab{}) - 2 \Lambda_2$ for the graviton,
and $\Deltab{} + \mathfrak R(\Deltab{})$ for the ghost.
As a downside, it doesn't regularize the pure gauge modes at all.
This poses no obstruction for the calculation of the flow of the fluctuation couplings.
As it turns out, their contribution to the flow of fluctuation couplings drops out in the Landau gauge
even for more general regulators which regularize all modes (see below).
Note however that this is not true for the flow of the background couplings.
There the above regulator is ill-behaving since it destroys the on-shell cancellation of the Faddeev-Popov ghosts
and the gauge modes, \ie{} one expects for the background cosmological constant a positive contribution
from five transverse traceless (TT) modes, three transverse vector modes and two 
scalar modes, and a negative contribution from the eight ghost modes, leading to two physical modes for the graviton.
However, this cancellation in the background sector is broken for the above regulator.
On the other hand, for fluctuation 
flows, this regulator has several virtues. First of all, it doesn't involve the gauge parameter $\alpha$, thus it allows for technical simplifications in the 
Landau limit, which can be taken already after the propagator has been calculated. Secondly, this regulator can be trivially extended to higher derivative 
theories.

A second possibility to regularize the propagator is to generalize the regulator of
\cite{Christiansen:2012rx, Christiansen:2014raa, Christiansen:2015rva, Denz:2016qks, Christiansen:2016sjn} to curved space.
For this to give well-defined flows, the ordering of the derivatives and the shape function is crucial.
To see this, first note that $[ f(\bar{\Delta}) , \bar{D}_{\mu} ] \neq 0$, since we are considering a curved space.
Second, there is a subtle interplay of taking the Landau limit and the propagator becoming degenerate in the
scalar sector of the TT-decomposition.
These two properties can lead to divergences, cf.\ equation (25) of \cite{Gies:2015tca}.
It turns out that the following ordering is well-defined, at least to linear order in the background curvature that we consider here:
\begin{widetext}
\begin{equation}\label{eq:reg_nontrivial}
\begin{aligned}
 \Delta S_h &= \frac{Z_h}{64\pi} \int \sqrt{\bar g} \, h_{\mu\nu} \Big[ -{\mathbbm 1^{\mu\nu}}_{\rho\sigma} \bar D^\alpha \mathfrak r(\Deltab{}) \bar D_\alpha 
- \frac{-8\alpha + (1+\beta)^2}{2\alpha} {\Pi_\text{Tr}^{\mu\nu}}_{\rho\sigma} \bar D^\alpha \mathfrak r(\Deltab{}) \bar D_\alpha \\
&\qquad\qquad\qquad\qquad  + \frac{1-2\alpha+\beta}{2} \Big( \bar g^{\mu\nu} \bar D_{(\rho} \mathfrak r(\Deltab{}) \bar D_{\sigma)}+\bar 
g_{\rho\sigma} \bar D^{(\mu} \mathfrak r(\Deltab{}) \bar D^{\nu)} \Big) + \delta^{(\mu}_{(\rho} \bar D^{\nu)} \mathfrak r(\Deltab{}) \bar D_{\sigma)} \Big] 
h^{\rho\sigma} \, .
\end{aligned}
\end{equation}
\end{widetext}
In this, we substituted $\mathfrak R(\Deltab{})$ by $\Deltab{} \mathfrak r(\Deltab{})$. In a similar way, for the ghosts we choose
\begin{equation}
 \Delta S_c = - \int \sqrt{\bar g} \, \bar c_\mu \left[ \delta^\mu_\nu \bar D_\rho \mathfrak r(\Deltab{}) \bar D^\rho + \frac{1-\beta}{2} \bar D^\mu \mathfrak 
r(\Deltab{}) \bar D_\nu \right] c^\nu \, .
\end{equation}
It is this regulator that we use in presenting numerical results. Still, as already stressed earlier, for the fluctuation flows on flat background, both 
regulators give the same result for the graviton contribution to the flow, which is a highly nontrivial result. Although small, a natural difference appears 
due to the different regularizations of the ghost modes. For the 1st order curvature couplings, also only minor differences in the flow induced by the 
gravitons arise.
For the actual shape function, we take the Litim regulator \cite{Litim:2001up},
\begin{equation}
 \mathfrak r(\bar{\Delta}) = \left( \frac{k^2}{\Deltab{}} - 1 \right) \theta\left(1 - \frac{\bar{\Delta}}{k^2} \right) \, ,
\end{equation}
where $\theta$ is the Heaviside theta function. 

\subsection{Projection of flow equations and Landau limit}\label{subsec:proj}

We now give some information on the projection scheme to extract beta functions, and comment on the Landau gauge. For the two-point function, all terms 
appearing on the right-hand side are sorted such that they have a similar form as \eqref{eq:action_quad}. From this we immediately get the flows of the first 
order
curvature couplings and the gaps. We can also extract the flow of the gauge parameters in this way, but it turns out that in the Landau limit no nontrivial 
contribution from their flow enters the flow equation of any other coupling. To make this precise, let us introduce the anomalous dimension
\begin{equation}
 \eta = - k \partial_k \ln Z_h \, ,
\end{equation}
and ``gauge anomalous dimensions'', cf.\ equation (3.18) of \cite{Benedetti:2011ct},
\begin{equation}
\begin{aligned}
 \eta_\xi &= -k \partial_k \ln Z_\xi \, ,
 \qquad Z_{\xi} = \frac{Z_{h}}{\alpha} \, ,
 \\
 \eta_{\bar{\sigma}} &= -k \partial_k \ln Z_{\bar{\sigma}} \, ,
 \qquad Z_{\bar{\sigma}} = \frac{Z_{h} (3-\beta)^{2}}{16 \alpha} \, .
\end{aligned}
\end{equation}
By direct evaluation, we get for the left-hand side of the flow equation,
\begin{equation}\label{eq:Flow_LHS}
\begin{aligned}
 k \partial_{k} Z_{\xi} &= - \frac{Z_{h}}{\alpha} \eta - \frac{Z_{h}}{\alpha^{2}} \dot{\alpha}
 \, ,
 \\
  k \partial_{k} Z_{\bar{\sigma}} &=
   - \frac{(3-\beta)^{2} Z_{h}}{16 \alpha} \eta
   - \frac{(3-\beta)^{2} Z_{h}}{16 \alpha^{2}} \dot{\alpha}
   - \frac{(3-\beta) Z_{h}}{8 \alpha} \dot{\beta} \, .
\end{aligned}
\end{equation}
On the other hand, the right-hand side of the flow equation evaluates to
\begin{equation}\label{eq:Flow_RHS}
\begin{aligned}
 k \partial_{k} Z_{\xi} &= \mcC_1 \dot{\alpha} + \mcC_2 \dot{\beta} + \mcC_3,
 \\
 k \partial_{k} Z_{\bar{\sigma}} &= \mcC_4 \dot{\alpha} + \mcC_5 \dot{\beta} + \mcC_6\, ,
\end{aligned}
\end{equation}
where the $\mcC_{i}$ are functions of the couplings and gauge parameters.
It is important to note that these stay finite in the Landau limit only if a well-behaved regulator is chosen. This is the case for both \eqref{eq:reg_minimal} 
and \eqref{eq:reg_nontrivial}, within our truncation.

Now we can equate both sides of the flow equation, i.e.\ we take \eqref{eq:Flow_RHS} and subtract
from this \eqref{eq:Flow_LHS}, to get
\begin{equation}
\begin{aligned}
  0
  ={}& \left( \frac{Z_{h}}{\alpha^{2}} + \mcC_{1} \right) \dot{\alpha}
  + \frac{Z_{h}}{\alpha} \eta + \mcC_{2} \dot{\beta} + \mcC_{3},
  \\
  0
  ={}& \left( \frac{(3-\beta)^{2} Z_{h}}{16 \alpha^{2}} + \mcC_{4} \right) \dot{\alpha}
  + \frac{(3-\beta)^{2} Z_{h}}{16 \alpha} \eta
  \\ {}&
  + \left( \frac{(3-\beta) Z_{h}}{8 \alpha} + \mcC_{5} \right) \dot{\beta}
  + \mcC_{6}.
\end{aligned}
\end{equation}
These equations can now be easily solved for $\dot{\alpha}$ and $\dot{\beta}$, but already
by inspection of the leading $\alpha \to 0$ divergences we find
\begin{equation}
\begin{aligned}
 \dot{\alpha} &= - \alpha \, \eta + \mcO(\alpha^{2}),
 \\
 \dot{\beta} &= - \alpha \frac{16 \, \mathcal C_6 - \mathcal C_3(3-\beta)^2}{2Z_h (3-\beta)} + \mcO(\alpha^{2}).
\end{aligned}
\end{equation}
Hence, we immediately infer that the Landau gauge is a fixed point
for both gauge parameters with arbitrary $\beta < 3$.
Equally well we can formulate this in terms of the gauge anomalous dimensions, $\eta_{\xi}$ and $\eta_{\bar{\sigma}}$,
\begin{equation}\label{eq:etaxi}
 \eta_{\xi} = \frac{\dot \alpha}{\alpha} + \eta = \mcO(\alpha) \, ,
 \qquad
 \eta_{\bar{\sigma}} = 2 \frac{\dot \beta}{3 - \beta} + \eta_{\xi} = \mcO(\alpha) \,.
\end{equation}

These results suggest that the gauge modes should not be rescaled by the wave function renormalization $Z_{h}$.
This removes the appearances of $Z_h$ and $\eta$ in the above equations,
and implies that both gauge parameters are exactly marginal,
since the stability matrix evaluated in the Landau limit has zero eigenvalues,
\begin{equation}
 \begin{pmatrix}
  \frac{\partial \dot\alpha}{\partial \alpha} & \frac{\partial \dot\alpha}{\partial \beta} \\
  \frac{\partial \dot\beta}{\partial \alpha} & \frac{\partial \dot\beta}{\partial \beta} \\
 \end{pmatrix}
 =
 \begin{pmatrix}
  0 & 0 \\
  \lim\limits_{\alpha\to0} \frac{3-\beta}{2\alpha} \left( \eta_{\bar\sigma} - \eta_\xi \right) & 0 \\
 \end{pmatrix} + \mathcal O(\alpha) \, .
\end{equation}
The limit is finite since both gauge anomalous dimensions are of $\mathcal O(\alpha)$, see \eqref{eq:etaxi}.

In the remainder of this work, we approximate $\eta=0$, which was shown to be a very good effective approximation of the fully momentum-dependent 
anomalous dimension \cite{Christiansen:2014raa}.

\begin{figure}
 \includegraphics[width=\columnwidth]{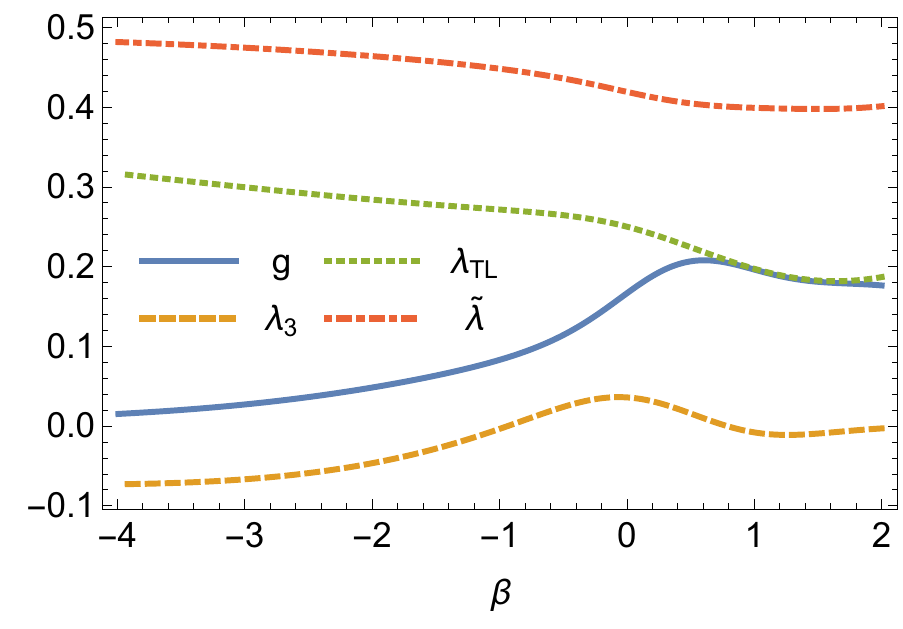}
 \caption{Dependence of the fixed point values of the 0th order curvature couplings on the gauge parameter $\beta$.
 All couplings show only a mild variation with $\beta$.
 In particular, both gaps are stable, which is to be contrasted with $\lambda_\text{Tr}$,
 which depends approximately quadratically on $\beta$, cf.\ equation \eqref{eq:lambdatilde}.}
 \label{fig:FPvals_flat}
\end{figure}

Finally, we have to specify the projection scheme for the couplings $G_3$ and $\Lambda_{3}$.
A full characterization of the three-point function seems difficult at present due to the high number of different
operators, see appendix \ref{app:basis}.
To enable checks with previous works, we choose the projection as in \cite{Christiansen:2015rva}.
In the language of appendix \ref{app:basis}, this amounts to projecting on the following linear combinations:
\begin{equation}
 \dot G_3 \sim -\frac{2}{3} c_{17} + \frac{1}{21} c_{13} - \frac{6}{21} c_{12} \, ,
 \qquad
 \dot \Lambda_3 \sim c_{3} \, .
\end{equation}
Note that the corresponding operators are exactly the ones containing the TT-mode of the fluctuation only. We verified that we get the 
same flow equations for $G_3$ and $\Lambda_3$ as given in \cite{Christiansen:2015rva} if we choose $\beta=1$ and identify $\Lambda_\text{Tr} = 
\Lambda_\text{TL} = \Lambda_2$. The Landau limit flow equations for all couplings are given in the supplemented notebook. To derive the flow equations, we used 
the Mathematica suite \textit{xAct} \cite{xActwebpage, Brizuela:2008ra, 
2008CoPhC.179..597M, 2007CoPhC.177..640M, 2008CoPhC.179..586M, 2014CoPhC.185.1719N}, and to calculate the traces we used covariant heat 
kernel techniques \cite{Barvinsky:1985an, Decanini:2005gt, Anselmi:2007eq, Groh:2011vn, Groh:2011dw}.

\section{Fixed point analysis}\label{sec:results}

We can now discuss the fixed point structure of our system. For this, we introduce dimensionless couplings in the following way:
\begin{equation}
\begin{aligned}
 g &= G_3 k^2 \, , \qquad \qquad & \lambda_3 &= \Lambda_3 / k^2 \, , \\
 \lambda_\text{TL} &= \Lambda_\text{TL} / k^2 \, , \qquad \qquad & \tilde\lambda &= \tilde\Lambda / k^2 \, .
\end{aligned}
\end{equation}
The 1st order curvature couplings are already dimensionless.

\begin{figure}
 \includegraphics[width=\columnwidth]{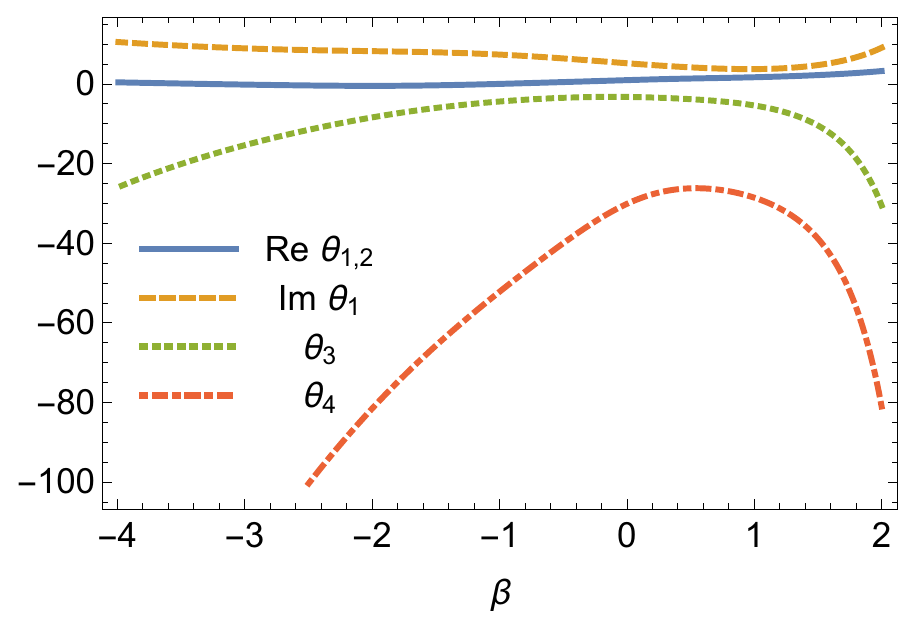}
 \caption{Dependence of the critical exponents of the 0th order curvature couplings on the gauge parameter $\beta$. The complex pair of critical exponents 
mainly 
corresponds to the couplings $g$ and $\lambda_\text{TL}$, the third mainly to $\lambda_3$,
and the most irrelevant one to $\tilde\lambda$.}
 \label{fig:CritExp_flat}
\end{figure}

First we discuss the flow of the couplings which also exist on a flat background, $(g, \lambda_\text{TL}, \tilde\lambda, \lambda_3)$, in the Landau limit, for 
arbitrary $\beta$. If these couplings don't show a fixed point, the full system cannot show it,
since the flow of these couplings by construction doesn't depend on 
the 1st order curvature couplings. Afterwards, we discuss the fate of the latter. For definiteness, we only discuss the results obtained with the regulator 
\eqref{eq:reg_nontrivial}. The results for the other regulator \eqref{eq:reg_minimal} are quantitatively very similar.

\subsection{0th order curvature couplings}\label{subsec:flat}

It turns out that we find a single fixed point which is rather stable under variation of the gauge parameter $\beta$. The fixed point values of the 0th order 
curvature couplings in dependence on $\beta$ are shown in \autoref{fig:FPvals_flat}. It can be seen that all couplings depend mildly on $\beta$. In particular, 
both gaps behave very similarly, and are effectively only shifted by a constant. \autoref{fig:CritExp_flat} shows the critical exponents of these couplings, 
being minus the eigenvalues of the stability matrix. We generically find one complex conjugate pair, and two real repulsive exponents. The complex pair 
corresponds to relevant operators for $\beta \gtrsim -0.98$, and to irrelevant operators for values of $\beta$ less than this. They have their main direction 
along the $(g,\lambda_\text{TL})$-plane. One should however note that in any case, the imaginary part dominates and the absolute value seems to be rather large,
indicating that further operators might be necessary to pin down the relevance of these operators. The generically irrelevant operators are $\lambda_3$ and 
$\tilde\lambda$, where the latter is more strongly irrelevant.

Typically, for a given value of $\beta$, other fixed points exist in the physical regime. As an example, for both choices $\beta=1$ and 
$\beta=-1$, we find a fixed point with two or one relevant direction, respectively. Nevertheless, changing $\beta$ reveals that these fixed points depend 
strongly on the gauge. This emphasizes the need to check gauge dependence if one wants to reliably select a suitable fixed point for the \UV{} completion of 
quantum gravity in an \RG{} setup.

Let us also note that in the limit $\beta\to-\infty$, which was preferred in the background field approximation \cite{Gies:2015tca} due to its weak gauge 
dependence, we don't find a physically interesting fixed point. This means that all fixed points that are found have either a negative $g$ or are behind the 
singularities at $\lambda_\text{TL} = 1/2$ and $\lambda_\text{TL} = -1/4$. The latter pole comes from the fact that the gap of the scalar modes,
$\tilde{\Lambda}$, given by 
\eqref{eq:lambdatilde}, doesn't include $\lambda_\text{Tr}$ in that limit,
\begin{align}
 \tilde{\Lambda} \stackrel{\beta \to - \infty}{\longrightarrow} - 2 \Lambda_{\text{TL}} \, ,
\end{align}
and thus there is a second pole induced solely by $\lambda_\text{TL}$.

\subsection{1st order curvature couplings}\label{subsec:curved}

\begin{figure}
 \includegraphics[width=\columnwidth]{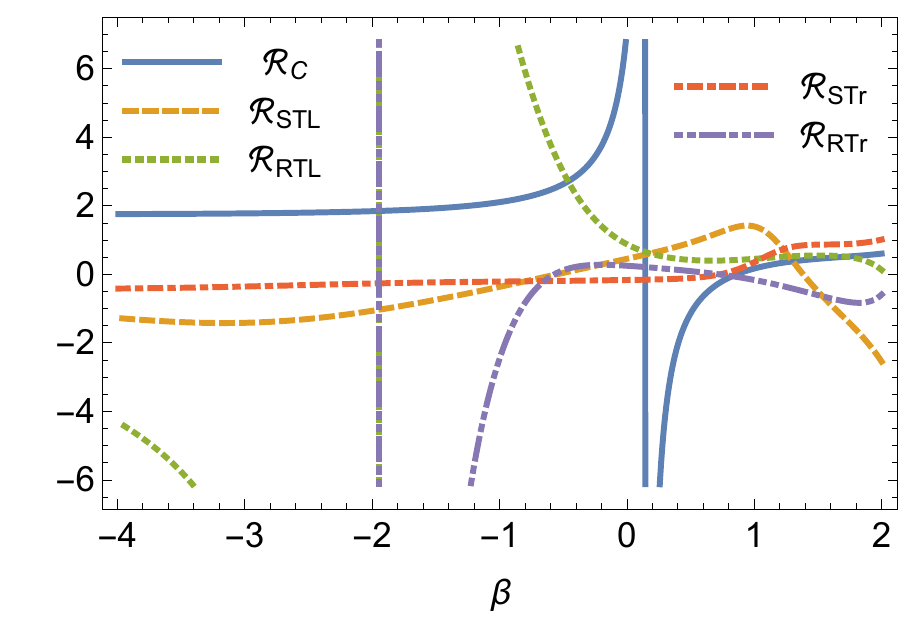}
 \caption{Dependence of the fixed point values of the 1st order curvature couplings on the gauge parameter $\beta$. Some couplings show divergences for 
specific values of $\beta$, indicating the breakdown of the truncation or the regulator.}
 \label{fig:FPvals_curvature}
\end{figure}

Let us now turn our attention to the 1st order curvature couplings. The fixed point values are shown in \autoref{fig:FPvals_curvature}. One can see a much 
stronger gauge dependence than in the case of the 0th order curvature couplings.
This has mainly two reasons. On the one hand, many further operators contribute to their flow equation, which however are higher order in our ordering scheme, 
\eg{} $h^{\mu\nu} \Rb{} \Deltab{} h_{\mu\nu}$. On the other hand, the regulator choice, in particular the inclusion of endomorphisms, decides how modes are 
integrated out, and thus it also has a leading order effect on the flow of the 1st order curvature couplings. It is hence not that surprising that some 
couplings even show divergences for specific choices of the gauge parameter $\beta$.

The 1st order curvature couplings are introduced in such a way that their classical value is zero, cf.\ \eqref{eq:action_quad}. Away from the singular points, 
their fixed point values are generically small, \ie{} of order one. This indicates that the quantum deviations of the diffeomorphism symmetry are of 
semi-perturbative nature. Note also that a comparison with background flows is inherently difficult, since our fluctuation setup comprises five different 
couplings belonging to linear order in the background curvature, whereas in a background setup, there is only one.

\begin{figure}
 \includegraphics[width=\columnwidth]{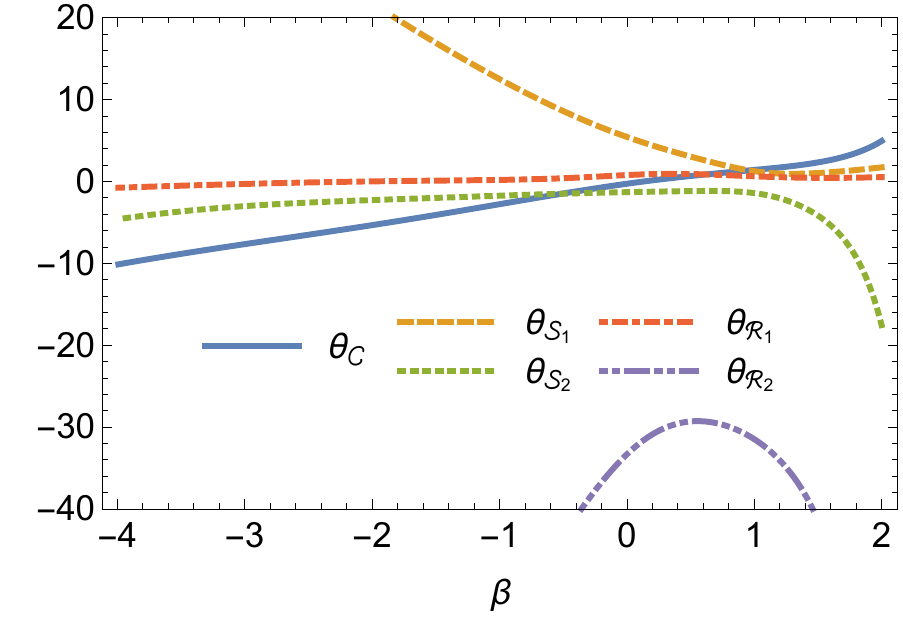}
 \caption{Dependence of the critical exponents of the 1st order curvature couplings on the gauge parameter $\beta$. They can be grouped according to the 
curvature tensor 
they refer to, since in the traceless basis different curvature tensors don't mix. Thus, $\theta_\mathcal C$ is exactly the critical exponent corresponding to 
$\mathcal R_\text{C}$, $\theta_{\mathcal S_{1,2}}$ corresponds to the mixing of $\mathcal R_\text{STL}$ and $\mathcal R_\text{STr}$, whereas 
$\theta_{\mathcal 
R_{1,2}}$ corresponds to the mixing of $\mathcal R_\text{RTL}$ and $\mathcal R_\text{RTr}$. At the values of $\beta$ where one of the couplings shows a 
divergence, one of the critical exponents changes sign, thus changing the relevance of the corresponding operator.}
 \label{fig:CritExp_curvature}
\end{figure}

In \autoref{fig:CritExp_curvature} we display the corresponding critical exponents. Due to the traceless 
basis we employ, the critical exponents cluster. This means that the eigenvectors of the critical exponents corresponding to the Weyl tensor invariant, the 
traceless Ricci tensor invariants and the Ricci scalar invariants are strictly orthogonal. In the mixing sectors, the more relevant operators are mainly 
corresponding to the couplings $\mathcal R_\text{STr}$ and $\mathcal R_\text{RTL}$. This is partly unexpected, since $\mathcal R_\text{STL}$ involves two 
external TT legs, whereas $\mathcal R_\text{STr}$ couples a TT leg and a Tr leg. Naively, this suggests that $\mathcal R_\text{STL}$ should be more relevant 
than $\mathcal R_\text{STr}$. By contrast, this expected ordering of relevance emerges for the operators involving the Ricci scalar.

The general trend is that two to three critical exponents are positive and thus correspond to relevant operators.
A comparatively strong gauge dependence indicates that either further operators need to be included, or endomorphisms have to be added in the 
regularization, to make conclusive statements.
Nevertheless, also irrelevant operators appear, which is encouraging for the Asymptotic Safety scenario.

\section{Conclusions}\label{sec:summary}

In this work we made progress on several frontiers of the Asymptotic Safety program for quantum gravity.
Above all, for the first time we resolved fluctuation correlation functions
on a generically curved background. For this, we studied the propagator to linear order in the background curvature. We further disentangled 
the flow of the two gaps of the graviton propagator. Together with a gauge-dependent redefinition of the scalar gap, we found a \UV{} fixed point suitable for 
Asymptotic Safety, where couplings vary only mildly for different gauge choices. The 1st order curvature couplings show a stronger gauge dependence, which is 
expected since they are of higher order, and further lower order operators, as $h^{\mu\nu}\Deltab{}^2h_{\mu\nu}$, will contribute to the leading order of their 
flow equations. Also, the choice of endomorphisms in the regularization plays an important role.

Furthermore we were able to explicitly check that the Landau gauge generically is a fixed point. Moreover, in this gauge also the gauge parameter $\beta$ does 
not flow. This behaviour only arises for a sensible choice of the regulator. Here we provided two examples of regularizations which are well-behaved to linear 
order in the background curvature, and distinguished by different tensor structures. Both regulators lead to agreeing results
for the graviton contribution to the flow of the 0th order curvature couplings.

Future work can go in several directions. For once, higher order correlation functions, as in \cite{Denz:2016qks}, should be included to stabilize the system. 
As a long term goal, all operators of the basis presented in the appendix \ref{app:basis} should be resolved. Compared to the present setting, this would 
include about 25 more couplings. One might also expect that momentum dependencies will become more important if the coupling to the background curvature is 
studied more extensively, since momentum and curvature are intimately related.

\section*{Acknowledgements}

We would like to thank H.\ Gies, J.\ M.\ Pawlowski and M.\ Reichert
for insightful discussions and H.\ Gies and J.\ M.\ Pawlowski
for valuable comments on the manuscript.
This work was supported by the Deutsche Forschungsgemeinschaft (DFG) Research Training Group ``Quantum and Gravitational Fields'' GRK 1523/2.
B.K.\ acknowledges funding by the DFG under grant no.\ Wi777/11.
S.L.\ acknowledges support by the DFG under grant no.\ Gi328/7-1.

\appendix

\section{Basis for correlation functions}\label{app:basis}

Here, we specify a basis for all correlation functions with up to three gravitons, including
up to two derivatives or one background curvature.
The one-point correlator has 3 independent structures,
\begin{equation}
 \Gamma^{(1)} \sim a_1 h + a_2 h \, \Rb{} + a_3 h_{\mu\nu} \Sb{}^{\mu\nu} \, .
\end{equation}
For the two-point function, there are 2 invariants without derivatives or curvature,
\begin{equation}
 \Gamma^{(2)}_\lambda \sim b_1 h^2 + b_2 h_{\mu\nu} h^{\mu\nu}\, ,
\end{equation}
4 invariants with 2 derivatives,
\begin{equation}
\begin{aligned}
 \Gamma^{(2)}_\text{D} &\sim b_3 h \, \Deltab{} \, h + b_4 h_{\mu\nu} \Deltab{} \, h^{\mu\nu} \\
 &\quad + b_5 h \Db{}^\mu \Db{}^\nu h_{\mu\nu} + b_6 \Db{}^\mu h_{\mu\rho} \Db{}^\nu {h_\nu}^\rho \, ,
\end{aligned}
\end{equation}
and 5 invariants with a background curvature,
\begin{equation}
\begin{aligned}
 \Gamma^{(2)}_\text{R} &\sim b_7 h_{\mu\nu} \Cb{}^{\mu\rho\nu\sigma} h_{\rho\sigma} + b_8 h_{\mu\nu} \Sb{}^{\mu\rho} {h_{\rho}}^\nu \\
 &\quad + b_9 h_{\mu\nu} \Sb{}^{\mu\nu} h + b_{10} h_{\mu\nu} \Rb{} \, h^{\mu\nu} + b_{11} h \, \Rb{} \, h \, .
\end{aligned}
\end{equation}
Finally, the three-point correlator can be spanned by 3 terms without derivatives or curvature,
\begin{equation}
 \Gamma^{(3)}_\lambda \sim c_1 h^3 + c_2 h_{\mu\nu} h^{\mu\nu} h + c_3 {h_\mu}^\nu {h_\nu}^\rho {h_\rho}^\mu \, ,
\end{equation}
14 terms with 2 derivatives,
\begin{widetext}
\begin{equation}
\begin{aligned}
 \Gamma^{(3)}_\text{D} &\sim c_4 h^2 \, \Deltab{} \, h + c_5 h^2 \Db{}_\mu \Db{}_\nu h^{\mu\nu} + c_6 h \, h^{\mu\nu} \Db{}_\mu \Db{}_\nu h + c_7 
{h_\mu}^\rho h^{\mu\sigma} \Db{}_{(\rho} \Db{}_{\sigma)} h + c_8 h \, \left( \Db{}_\rho {h_\mu}^\rho \right) \Db{}_\sigma h^{\mu\sigma} \\
 &\quad + c_9 h \, {h_\mu}^\rho \Db{}_{(\rho} \Db{}_{\sigma)} h^{\mu\sigma} + c_{10} h \, h_{\mu\nu} \Deltab{} \, h^{\mu\nu} + c_{11} h_{\mu\nu} 
h^{\mu\nu} \Deltab{} \, h + c_{12} {h_\mu}^\rho {h_\rho}^\nu \Deltab{} \, {h_\nu}^\mu + c_{13} h_{\mu\nu} h_{\rho\sigma} \Db{}^\mu \Db{}^\nu h^{\rho\sigma} \\
 &\quad + c_{14} h_{\mu\nu} h^{\mu\nu} \Db{}_\rho \Db{}_\sigma h^{\rho\sigma} + c_{15} h^{\rho\nu} h_{\sigma\nu} \Db{}_{(\mu} \Db{}_{\rho)} 
h^{\mu\sigma} + c_{16} h_{\nu\sigma} \left( \Db{}_\mu h^{\mu\nu} \right) \Db{}_\rho h^{\rho\sigma} + c_{17} h^{\mu\nu} h^{\rho\sigma} \Db{}_{(\mu} 
\Db{}_{\rho)} h_{\nu\sigma} \, ,
\end{aligned}
\end{equation}
and 9 invariants with a background curvature,
\begin{equation}
\begin{aligned}
 \Gamma^{(3)}_\text{R} &\sim c_{18} \Cb{}^{\mu\nu\rho\sigma} h_{\nu\sigma} h_{\mu\tau} {h_\rho}^\tau + c_{19} \Cb{}^{\mu\nu\rho\sigma} h \, h_{\mu\rho} 
h_{\nu\sigma} + c_{20} \Sb{}^{\mu\nu} {h_\nu}^\rho {h_\rho}^\sigma h_{\sigma\mu} + c_{21} \Sb{}^{\mu\nu} h_{\mu\nu} h^{\rho\sigma} h_{\rho\sigma} \\
 &\quad + c_{22} \Sb{}^{\mu\nu} {h_\nu}^\rho {h_\rho}^\mu h + c_{23} \Sb{}^{\mu\nu} h_{\mu\nu} h^2 + c_{24} \Rb{} \, {h_\mu}^\nu {h_\nu}^\rho 
{h_\rho}^\mu + c_{25} \Rb{} \, h_{\mu\nu} h^{\mu\nu} h + c_{26} \Rb{} \, h^3 \, .
\end{aligned}
\end{equation}
\end{widetext}

\section{General formula for the propagator on flat background}\label{app:prop}

In this section, we show how to get the propagator on a flat background, for a general ansatz independent of the truncation.
The most general form of a symmetric rank (2,2) tensor $T$ on a flat background depending on a single momentum vector $p$ reads
\begin{equation}
\begin{aligned}
 {T^{\mu\nu}}_{\rho\sigma} &= A_1 \tensor{{\Pi_{\mathrm{TL}}}}{^{\mu\nu}_{\rho\sigma}} + A_2 \, \tensor{{\Pi_{\mathrm{Tr}}}}{^{\mu\nu}_{\rho\sigma}}
 + A_3 \, p^{(\mu} \delta^{\nu)}_{(\rho} p_{\sigma)}
 \\
 &\quad  + \tfrac{A_4}{2} \left( p^\mu p^\nu \gb{}_{\rho\sigma} + \gb{}^{\mu\nu} p_\rho p_\sigma \right)
 + A_5 \, p^\mu p^\nu p_\rho p_\sigma \, .
\end{aligned}
\end{equation}
Both the second variation of the action and the propagator are of this form.
It is straightforward to calculate the inverse of this tensor,
\begin{equation}
 {T^{\mu\nu}}_{\rho\sigma} {{\left(T^{-1}\right)}^{\rho\sigma}}_{\alpha\beta} = {\mathbbm{1}^{\mu\nu}}_{\alpha\beta} \, ,
\end{equation}
by explicit insertion of the ansatz.
The coefficients $B_i$ of the inverse of the tensor with coefficients $A_i$ read
\begin{widetext}
\begin{equation}\label{eq:B2ARule}
\begin{aligned}
 B_1 &= \frac{1}{A_1} \, ,
 \\
 B_2 &= \frac{4 A_1^2 + \big( A_1 ( 5 A_3 + 4 A_4 ) - A_2 A_3 \big) p^2 + \big( (5 A_1 - A_2 ) A_5 + A_4^2 \big) p^4}%
 {4 A_1^2 A_2 + A_1 c_1 p^2 + A_1 c_2 p^4} \, ,
 \\
 B_3 &= -\frac{2 A_3}{ 2 A_1^2 + A_1 A_3 p^2} \, ,
 \\
 B_4 &= \frac{2 \big( A_1 ( - A_3 - 2 A_4 ) + A_2 A_3 \big) - 2 \big( (A_1 - A_2) A_5 + A_4^2 \big) p^2}%
 {4 A_1^2 A_2 + A_1 c_1 p^2 + A_1 c_2 p^4} \, ,
 \\
 B_5 &= \frac{2 \big( A_1 \big( (A_3 + 2 A_4)^{2} - 4 A_2 A_5 \big) + A_2 A_3^2 \big) + 2 A_3 ( (A_1 + A_2) A_5 - A_4^2 ) p^2}%
 {8 A_1^3 A_2 + 2 A_1^2 ( c_1 + 2 A_2 A_3 ) p^2
 + A_1 \big( 2 A_1 c_2 + A_3 c_1 \big) p^4
 + A_1 A_3 c_2 p^6} \, ,
\end{aligned}
\end{equation}
\end{widetext}
where $p^2 = p_\mu p^\mu$, $c_1 = A_1 (A_3 + 4 A_4) + 3 A_2 A_3$ and $c_2 = (A_1 + 3 A_2) A_5 - 3 A_4^2$.
For small $p^2$, only $A_1$ and $A_2$ appear in the denominators, thus any regulator has to regularize at least these two structures.
For computational reasons, we often need derivatives of the $B_{i}$ w.r.t.\ $p^2$.
As the expressions \eqref{eq:B2ARule} are rational functions, their derivatives are rather lengthy.
Here it helps to note, that for a simple function of the form $g(x) = \frac{1}{f(x)}$ the derivative can be written as $g'(x) = - g^{2}(x) f'(x)$.
Similar expressions for the derivatives of the $B_{i}$ can be found if we rescale the $A_{i}$ and the $B_{i}$ in such a way that no explicit $p^2$ appears in 
their relation,
\begin{widetext}
\begin{equation}
\begin{aligned}
 \tilde{A}_1 = A_1 \, ,
 \quad
 \tilde{A}_2 = A_2 \, ,
 \quad
 \tilde{A}_3 = p^2 A_3 \, ,
 \quad
 \tilde{A}_4 = p^2 A_4 \, ,
 \quad
 \tilde{A}_5 = p^4 A_5 \, , \\
 \tilde{B}_1 = B_1 \, ,
 \quad
 \tilde{B}_2 = B_2 \, ,
 \quad
 \tilde{B}_3 = p^2 B_3 \, ,
 \quad
 \tilde{B}_4 = p^2 B_4 \, ,
 \quad
 \tilde{B}_5 = p^4 B_5 \, .
\end{aligned}
\end{equation}
Then the $\tilde{B}_{i}$ as a function of the $\tilde{A}_{i}$ are given by equations \eqref{eq:B2ARule},
replacing $A_{i} \to \tilde{A}_{i}$, $B_{i} \to \tilde{B}_{i}$ and $p^2 \to 1$.
Then one can check that the derivatives of the $\tilde{B}_{i}$ are given by
\begin{equation}
\begin{aligned}
 \tilde{B}'_{1} ={}& - \tilde{B}_{1}^2 \cdot \tilde{A}_{1}' \, ,
\\
 \tilde{B}'_{2} ={}& - \tfrac{1}{4} ( - 4 \tilde{B}_1 + 3 \tilde{B}_4 ) \tilde{B}_4 \cdot \tilde{A}'_1 
 - \tfrac{1}{4} ( 2 \tilde{B}_2 + \tilde{B}_4 )^{2} \cdot \tilde{A}'_2
 - \tfrac{1}{4} d_1^{2} \cdot \tilde{A}'_3
 - \tfrac{1}{2} d_1 ( 2 \tilde{B}_2 + \tilde{B}_4 ) \cdot \tilde{A}'_4
 - \tfrac{1}{4} d_1^{2} \cdot \tilde{A}'_5 \, ,
\\
 \tilde{B}'_{3} ={}& - \tfrac{1}{2} ( 4 \tilde{B}_1 + \tilde{B}_3 ) \tilde{B}_3 \cdot \tilde{A}'_1 
 - \tfrac{1}{4} ( 2 \tilde{B}_1 + \tilde{B}_3 )^{2} \cdot \tilde{A}'_3 \, ,
\\
 \tilde{B}'_{4} ={}& - \tfrac{1}{4} \big( 4 \tilde{B}_1 \tilde{B}_4 - ( 2 \tilde{B}_1 - 3 \tilde{B}_4 ) ( \tilde{B}_3 + \tilde{B}_5 ) \big) \cdot \tilde{A}'_1 
 - \tfrac{1}{4} ( 2 \tilde{B}_2 + \tilde{B}_4 ) ( \tilde{B}_3 + 2 \tilde{B}_4 + \tilde{B}_5 ) \cdot \tilde{A}'_2
\\
 {}& - \tfrac{1}{4} d_1 d_2 \cdot \tilde{A}'_3
 - \tfrac{1}{4} \big( ( - \tilde{B}_1 + 5 \tilde{B}_2 + 4 \tilde{B}_4  ) ( \tilde{B}_3 + \tilde{B}_5 ) + 4 \tilde{B}_1 \tilde{B}_2 + 5 \tilde{B}_4^2 + 4 \tilde{B}_2 \tilde{B}_4 \big) \cdot \tilde{A}'_4
 - \tfrac{1}{4} d_1 d_2 \cdot \tilde{A}'_5 \, ,
\\
 \tilde{B}'_{5} ={}& - \tfrac{1}{4} \big( ( 8 \tilde{B}_1 + 3 \tilde{B}_5 ) \tilde{B}_5 + 6 \tilde{B}_3 \tilde{B}_5 + \tilde{B}_3^2 \big) \cdot \tilde{A}'_1 
 - \tfrac{1}{4} ( \tilde{B}_3 + 2 \tilde{B}_4 + \tilde{B}_5 )^{2} \cdot \tilde{A}'_2
\\
 {}& - \tfrac{1}{4} ( \tilde{B}_3 + \tilde{B}_4 + 2 \tilde{B}_5 ) ( 4 \tilde{B}_1 + 3 \tilde{B}_3 + \tilde{B}_4 + 2 \tilde{B}_5 ) \cdot \tilde{A}'_3
 - \tfrac{1}{2} d_2 ( \tilde{B}_3 + 2 \tilde{B}_4 + \tilde{B}_5 ) \cdot \tilde{A}'_4
 - \tfrac{1}{4} d_2^{2} \cdot \tilde{A}'_5
 \, ,
\end{aligned}
\end{equation}
where $d_1 = - \tilde{B}_1 + \tilde{B}_2 + 2 \tilde{B}_4$ and $d_2 = 2 \tilde{B}_1 + 2 \tilde{B}_3 + \tilde{B}_4 + 2 \tilde{B}_5$.
\end{widetext}

\section{Regulator comparison}\label{app:regcomp}

In this appendix, we present a comparison of fixed point values and critical exponents for both regulators.
In particular, we choose $\beta=1$, since close to this value all fixed point quantities seem to show a weak $\beta$-dependence, see the figures in the main 
text. Specifically, the most irrelevant critical exponents, $\theta_4$ and $\theta_{\mathcal R_2}$, which in general strongly depend on $\beta$, show a 
local maximum near to this value of $\beta$. Moreover, for this choice, both regulators agree completely on a flat background, and only differ 
to linear order in the background curvature. For the 0th order curvature couplings, we obtain
\begin{equation}
\begin{aligned}
 g &= 0.196 \, , \qquad \qquad &\lambda_3 &= -0.00807 \, , \\
 \lambda_\text{TL} &= 0.197 \, , \qquad \qquad &\tilde\lambda &= 0.399 \, ,
\end{aligned}
\end{equation}
together with the critical exponents
\begin{equation}
 \theta_{1,2} = 1.65 \pm 3.70 \mathbf{i} \, , \qquad \theta_3 = -5.43 \, , \qquad \theta_4 = -28.6 \, .
\end{equation}
For the 1st order curvature couplings, we find that only some of them have differing fixed point values due to the different regularization. With the minimal 
regulator \eqref{eq:reg_minimal}, we obtain
%
%
%
\begin{equation}
\begin{aligned}
 \mathcal R_\text{C} &= 0.164 \, , \qquad \qquad &\theta_C &= 1.39 \, , \\
 \mathcal R_\text{STL} &= 1.13 \, , \qquad \qquad &\theta_{S_1} &= 1.24 \, , \\
 \mathcal R_\text{STr} &= 0.476 \, , \qquad \qquad &\theta_{S_2} &= -1.44 \, , \\
 \mathcal R_\text{RTL} &= 0.453 \, , \qquad \qquad &\theta_{R_1} &= 0.607 \, , \\
 \mathcal R_\text{RTr} &= -0.252 \, , \qquad \qquad &\theta_{R_2} &= -31.5 \, .
\end{aligned}
\end{equation}
Employing the regulator \eqref{eq:reg_nontrivial}, we find
\begin{equation}
\begin{aligned}
 \mathcal R_\text{C} &= 0.164 \, , \qquad \qquad &\theta_C &= 1.39 \, , \\
 \mathcal R_\text{STL} &= 1.39 \, , \qquad \qquad &\theta_{S_1} &= 1.24 \, , \\
 \mathcal R_\text{STr} &= 0.348 \, , \qquad \qquad &\theta_{S_2} &= -1.44 \, , \\
 \mathcal R_\text{RTL} &= 0.447 \, , \qquad \qquad &\theta_{R_1} &= 0.607 \, , \\
 \mathcal R_\text{RTr} &= -0.169 \, , \qquad \qquad &\theta_{R_2} &= -31.5 \, .
\end{aligned}
\end{equation}
It can be seen that the fixed point value of $\mathcal R_\text{C}$ is the same for both regulators. Even more surprisingly, the critical exponents do not 
depend at all on the regulator for this choice of the gauge fixing. For general choices of $\beta$, there is a small difference between the two regulators in 
all couplings and critical exponents, typically on the percent level.

~

~

\bibliography{general_bib}

\begin{thebibliography}{141}%
\makeatletter
\providecommand \@ifxundefined [1]{%
 \@ifx{#1\undefined}
}%
\providecommand \@ifnum [1]{%
 \ifnum #1\expandafter \@firstoftwo
 \else \expandafter \@secondoftwo
 \fi
}%
\providecommand \@ifx [1]{%
 \ifx #1\expandafter \@firstoftwo
 \else \expandafter \@secondoftwo
 \fi
}%
\providecommand \natexlab [1]{#1}%
\providecommand \enquote  [1]{``#1''}%
\providecommand \bibnamefont  [1]{#1}%
\providecommand \bibfnamefont [1]{#1}%
\providecommand \citenamefont [1]{#1}%
\providecommand \href@noop [0]{\@secondoftwo}%
\providecommand \href [0]{\begingroup \@sanitize@url \@href}%
\providecommand \@href[1]{\@@startlink{#1}\@@href}%
\providecommand \@@href[1]{\endgroup#1\@@endlink}%
\providecommand \@sanitize@url [0]{\catcode `\\12\catcode `\$12\catcode
  `\&12\catcode `\#12\catcode `\^12\catcode `\_12\catcode `\%12\relax}%
\providecommand \@@startlink[1]{}%
\providecommand \@@endlink[0]{}%
\providecommand \url  [0]{\begingroup\@sanitize@url \@url }%
\providecommand \@url [1]{\endgroup\@href {#1}{\urlprefix }}%
\providecommand \urlprefix  [0]{URL }%
\providecommand \Eprint [0]{\href }%
\providecommand \doibase [0]{http://dx.doi.org/}%
\providecommand \selectlanguage [0]{\@gobble}%
\providecommand \bibinfo  [0]{\@secondoftwo}%
\providecommand \bibfield  [0]{\@secondoftwo}%
\providecommand \translation [1]{[#1]}%
\providecommand \BibitemOpen [0]{}%
\providecommand \bibitemStop [0]{}%
\providecommand \bibitemNoStop [0]{.\EOS\space}%
\providecommand \EOS [0]{\spacefactor3000\relax}%
\providecommand \BibitemShut  [1]{\csname bibitem#1\endcsname}%
\let\auto@bib@innerbib\@empty
\bibitem [{\citenamefont {Weinberg}(1979)}]{Weinberg:1980gg}%
  \BibitemOpen
  \bibfield  {author} {\bibinfo {author} {\bibfnamefont {S.}~\bibnamefont
  {Weinberg}},\ }\href@noop {} {\bibfield  {journal} {\bibinfo  {journal}
  {General Relativity: An Einstein centenary survey, Eds. Hawking, S.W.,
  Israel, W; Cambridge University Press}\ ,\ \bibinfo {pages} {790}} (\bibinfo
  {year} {1979})}\BibitemShut {NoStop}%
\bibitem [{\citenamefont {Wetterich}(1993)}]{Wetterich:1992yh}%
  \BibitemOpen
  \bibfield  {author} {\bibinfo {author} {\bibfnamefont {C.}~\bibnamefont
  {Wetterich}},\ }\href {\doibase 10.1016/0370-2693(93)90726-X} {\bibfield
  {journal} {\bibinfo  {journal} {Phys.Lett.}\ }\textbf {\bibinfo {volume}
  {B301}},\ \bibinfo {pages} {90} (\bibinfo {year} {1993})}\BibitemShut
  {NoStop}%
\bibitem [{\citenamefont {Ellwanger}(1994)}]{Ellwanger1994}%
  \BibitemOpen
  \bibfield  {author} {\bibinfo {author} {\bibfnamefont {U.}~\bibnamefont
  {Ellwanger}},\ }\href {\doibase 10.1007/BF01555911} {\bibfield  {journal}
  {\bibinfo  {journal} {Zeitschrift f{\"u}r Physik C Particles and Fields}\
  }\textbf {\bibinfo {volume} {62}},\ \bibinfo {pages} {503} (\bibinfo {year}
  {1994})}\BibitemShut {NoStop}%
\bibitem [{\citenamefont {Morris}(1994)}]{Morris:1993qb}%
  \BibitemOpen
  \bibfield  {author} {\bibinfo {author} {\bibfnamefont {T.~R.}\ \bibnamefont
  {Morris}},\ }\href {\doibase 10.1142/S0217751X94000972} {\bibfield  {journal}
  {\bibinfo  {journal} {Int. J. Mod. Phys.}\ }\textbf {\bibinfo {volume}
  {A9}},\ \bibinfo {pages} {2411} (\bibinfo {year} {1994})},\ \Eprint
  {http://arxiv.org/abs/hep-ph/9308265} {arXiv:hep-ph/9308265 [hep-ph]}
  \BibitemShut {NoStop}%
\bibitem [{\citenamefont {Reuter}(1998)}]{Reuter:1996cp}%
  \BibitemOpen
  \bibfield  {author} {\bibinfo {author} {\bibfnamefont {M.}~\bibnamefont
  {Reuter}},\ }\href {\doibase 10.1103/PhysRevD.57.971} {\bibfield  {journal}
  {\bibinfo  {journal} {Phys.Rev.}\ }\textbf {\bibinfo {volume} {D57}},\
  \bibinfo {pages} {971} (\bibinfo {year} {1998})},\ \Eprint
  {http://arxiv.org/abs/hep-th/9605030} {arXiv:hep-th/9605030 [hep-th]}
  \BibitemShut {NoStop}%
\bibitem [{\citenamefont {Gies}\ \emph {et~al.}(2015)\citenamefont {Gies},
  \citenamefont {Knorr},\ and\ \citenamefont {Lippoldt}}]{Gies:2015tca}%
  \BibitemOpen
  \bibfield  {author} {\bibinfo {author} {\bibfnamefont {H.}~\bibnamefont
  {Gies}}, \bibinfo {author} {\bibfnamefont {B.}~\bibnamefont {Knorr}}, \ and\
  \bibinfo {author} {\bibfnamefont {S.}~\bibnamefont {Lippoldt}},\ }\href
  {\doibase 10.1103/PhysRevD.92.084020} {\bibfield  {journal} {\bibinfo
  {journal} {Phys. Rev.}\ }\textbf {\bibinfo {volume} {D92}},\ \bibinfo {pages}
  {084020} (\bibinfo {year} {2015})},\ \Eprint
  {http://arxiv.org/abs/1507.08859} {arXiv:1507.08859 [hep-th]} \BibitemShut
  {NoStop}%
\bibitem [{\citenamefont {Souma}(1999)}]{Souma:1999at}%
  \BibitemOpen
  \bibfield  {author} {\bibinfo {author} {\bibfnamefont {W.}~\bibnamefont
  {Souma}},\ }\href {\doibase 10.1143/PTP.102.181} {\bibfield  {journal}
  {\bibinfo  {journal} {Prog. Theor. Phys.}\ }\textbf {\bibinfo {volume}
  {102}},\ \bibinfo {pages} {181} (\bibinfo {year} {1999})},\ \Eprint
  {http://arxiv.org/abs/hep-th/9907027} {arXiv:hep-th/9907027 [hep-th]}
  \BibitemShut {NoStop}%
\bibitem [{\citenamefont {Lauscher}\ and\ \citenamefont
  {Reuter}(2002{\natexlab{a}})}]{Lauscher:2001ya}%
  \BibitemOpen
  \bibfield  {author} {\bibinfo {author} {\bibfnamefont {O.}~\bibnamefont
  {Lauscher}}\ and\ \bibinfo {author} {\bibfnamefont {M.}~\bibnamefont
  {Reuter}},\ }\href {\doibase 10.1103/PhysRevD.65.025013} {\bibfield
  {journal} {\bibinfo  {journal} {Phys.Rev.}\ }\textbf {\bibinfo {volume}
  {D65}},\ \bibinfo {pages} {025013} (\bibinfo {year} {2002}{\natexlab{a}})},\
  \Eprint {http://arxiv.org/abs/hep-th/0108040} {arXiv:hep-th/0108040 [hep-th]}
  \BibitemShut {NoStop}%
\bibitem [{\citenamefont {Lauscher}\ and\ \citenamefont
  {Reuter}(2002{\natexlab{b}})}]{Lauscher:2001rz}%
  \BibitemOpen
  \bibfield  {author} {\bibinfo {author} {\bibfnamefont {O.}~\bibnamefont
  {Lauscher}}\ and\ \bibinfo {author} {\bibfnamefont {M.}~\bibnamefont
  {Reuter}},\ }\href {\doibase 10.1088/0264-9381/19/3/304} {\bibfield
  {journal} {\bibinfo  {journal} {Class. Quant. Grav.}\ }\textbf {\bibinfo
  {volume} {19}},\ \bibinfo {pages} {483} (\bibinfo {year}
  {2002}{\natexlab{b}})},\ \Eprint {http://arxiv.org/abs/hep-th/0110021}
  {arXiv:hep-th/0110021 [hep-th]} \BibitemShut {NoStop}%
\bibitem [{\citenamefont {Reuter}\ and\ \citenamefont
  {Saueressig}(2002)}]{Reuter:2001ag}%
  \BibitemOpen
  \bibfield  {author} {\bibinfo {author} {\bibfnamefont {M.}~\bibnamefont
  {Reuter}}\ and\ \bibinfo {author} {\bibfnamefont {F.}~\bibnamefont
  {Saueressig}},\ }\href {\doibase 10.1103/PhysRevD.65.065016} {\bibfield
  {journal} {\bibinfo  {journal} {Phys. Rev.}\ }\textbf {\bibinfo {volume}
  {D65}},\ \bibinfo {pages} {065016} (\bibinfo {year} {2002})},\ \Eprint
  {http://arxiv.org/abs/hep-th/0110054} {arXiv:hep-th/0110054 [hep-th]}
  \BibitemShut {NoStop}%
\bibitem [{\citenamefont {Litim}(2004)}]{Litim:2003vp}%
  \BibitemOpen
  \bibfield  {author} {\bibinfo {author} {\bibfnamefont {D.~F.}\ \bibnamefont
  {Litim}},\ }\href {\doibase 10.1103/PhysRevLett.92.201301} {\bibfield
  {journal} {\bibinfo  {journal} {Phys. Rev. Lett.}\ }\textbf {\bibinfo
  {volume} {92}},\ \bibinfo {pages} {201301} (\bibinfo {year} {2004})},\
  \Eprint {http://arxiv.org/abs/hep-th/0312114} {arXiv:hep-th/0312114 [hep-th]}
  \BibitemShut {NoStop}%
\bibitem [{\citenamefont {Lauscher}\ and\ \citenamefont
  {Reuter}(2005)}]{Lauscher:2005qz}%
  \BibitemOpen
  \bibfield  {author} {\bibinfo {author} {\bibfnamefont {O.}~\bibnamefont
  {Lauscher}}\ and\ \bibinfo {author} {\bibfnamefont {M.}~\bibnamefont
  {Reuter}},\ }\href {\doibase 10.1088/1126-6708/2005/10/050} {\bibfield
  {journal} {\bibinfo  {journal} {JHEP}\ }\textbf {\bibinfo {volume} {10}},\
  \bibinfo {pages} {050} (\bibinfo {year} {2005})},\ \Eprint
  {http://arxiv.org/abs/hep-th/0508202} {arXiv:hep-th/0508202 [hep-th]}
  \BibitemShut {NoStop}%
\bibitem [{\citenamefont {Reuter}\ and\ \citenamefont
  {Schwindt}(2006)}]{Reuter:2005bb}%
  \BibitemOpen
  \bibfield  {author} {\bibinfo {author} {\bibfnamefont {M.}~\bibnamefont
  {Reuter}}\ and\ \bibinfo {author} {\bibfnamefont {J.-M.}\ \bibnamefont
  {Schwindt}},\ }\href {\doibase 10.1088/1126-6708/2006/01/070} {\bibfield
  {journal} {\bibinfo  {journal} {JHEP}\ }\textbf {\bibinfo {volume} {01}},\
  \bibinfo {pages} {070} (\bibinfo {year} {2006})},\ \Eprint
  {http://arxiv.org/abs/hep-th/0511021} {arXiv:hep-th/0511021 [hep-th]}
  \BibitemShut {NoStop}%
\bibitem [{\citenamefont {Niedermaier}\ and\ \citenamefont
  {Reuter}(2006)}]{Niedermaier:2006wt}%
  \BibitemOpen
  \bibfield  {author} {\bibinfo {author} {\bibfnamefont {M.}~\bibnamefont
  {Niedermaier}}\ and\ \bibinfo {author} {\bibfnamefont {M.}~\bibnamefont
  {Reuter}},\ }\href {\doibase 10.12942/lrr-2006-5} {\bibfield  {journal}
  {\bibinfo  {journal} {Living Rev.Rel.}\ }\textbf {\bibinfo {volume} {9}},\
  \bibinfo {pages} {5} (\bibinfo {year} {2006})}\BibitemShut {NoStop}%
\bibitem [{\citenamefont {Groh}\ and\ \citenamefont
  {Saueressig}(2010)}]{Groh:2010ta}%
  \BibitemOpen
  \bibfield  {author} {\bibinfo {author} {\bibfnamefont {K.}~\bibnamefont
  {Groh}}\ and\ \bibinfo {author} {\bibfnamefont {F.}~\bibnamefont
  {Saueressig}},\ }\href {\doibase 10.1088/1751-8113/43/36/365403} {\bibfield
  {journal} {\bibinfo  {journal} {J. Phys.}\ }\textbf {\bibinfo {volume}
  {A43}},\ \bibinfo {pages} {365403} (\bibinfo {year} {2010})},\ \Eprint
  {http://arxiv.org/abs/1001.5032} {arXiv:1001.5032 [hep-th]} \BibitemShut
  {NoStop}%
\bibitem [{\citenamefont {Benedetti}\ \emph {et~al.}(2011)\citenamefont
  {Benedetti}, \citenamefont {Groh}, \citenamefont {Machado},\ and\
  \citenamefont {Saueressig}}]{Benedetti:2010nr}%
  \BibitemOpen
  \bibfield  {author} {\bibinfo {author} {\bibfnamefont {D.}~\bibnamefont
  {Benedetti}}, \bibinfo {author} {\bibfnamefont {K.}~\bibnamefont {Groh}},
  \bibinfo {author} {\bibfnamefont {P.~F.}\ \bibnamefont {Machado}}, \ and\
  \bibinfo {author} {\bibfnamefont {F.}~\bibnamefont {Saueressig}},\ }\href
  {\doibase 10.1007/JHEP06(2011)079} {\bibfield  {journal} {\bibinfo  {journal}
  {JHEP}\ }\textbf {\bibinfo {volume} {1106}},\ \bibinfo {pages} {079}
  (\bibinfo {year} {2011})},\ \Eprint {http://arxiv.org/abs/1012.3081}
  {arXiv:1012.3081 [hep-th]} \BibitemShut {NoStop}%
\bibitem [{\citenamefont {Manrique}\ \emph
  {et~al.}(2011{\natexlab{a}})\citenamefont {Manrique}, \citenamefont
  {Rechenberger},\ and\ \citenamefont {Saueressig}}]{Manrique:2011jc}%
  \BibitemOpen
  \bibfield  {author} {\bibinfo {author} {\bibfnamefont {E.}~\bibnamefont
  {Manrique}}, \bibinfo {author} {\bibfnamefont {S.}~\bibnamefont
  {Rechenberger}}, \ and\ \bibinfo {author} {\bibfnamefont {F.}~\bibnamefont
  {Saueressig}},\ }\href {\doibase 10.1103/PhysRevLett.106.251302} {\bibfield
  {journal} {\bibinfo  {journal} {Phys. Rev. Lett.}\ }\textbf {\bibinfo
  {volume} {106}},\ \bibinfo {pages} {251302} (\bibinfo {year}
  {2011}{\natexlab{a}})},\ \Eprint {http://arxiv.org/abs/1102.5012}
  {arXiv:1102.5012 [hep-th]} \BibitemShut {NoStop}%
\bibitem [{\citenamefont {Reuter}\ and\ \citenamefont
  {Saueressig}(2012)}]{Reuter:2012id}%
  \BibitemOpen
  \bibfield  {author} {\bibinfo {author} {\bibfnamefont {M.}~\bibnamefont
  {Reuter}}\ and\ \bibinfo {author} {\bibfnamefont {F.}~\bibnamefont
  {Saueressig}},\ }\href {\doibase 10.1088/1367-2630/14/5/055022} {\bibfield
  {journal} {\bibinfo  {journal} {New J.Phys.}\ }\textbf {\bibinfo {volume}
  {14}},\ \bibinfo {pages} {055022} (\bibinfo {year} {2012})},\ \Eprint
  {http://arxiv.org/abs/1202.2274} {arXiv:1202.2274 [hep-th]} \BibitemShut
  {NoStop}%
\bibitem [{\citenamefont {Harst}\ and\ \citenamefont
  {Reuter}(2012)}]{Harst:2012ni}%
  \BibitemOpen
  \bibfield  {author} {\bibinfo {author} {\bibfnamefont {U.}~\bibnamefont
  {Harst}}\ and\ \bibinfo {author} {\bibfnamefont {M.}~\bibnamefont {Reuter}},\
  }\href {\doibase 10.1007/JHEP05(2012)005} {\bibfield  {journal} {\bibinfo
  {journal} {JHEP}\ }\textbf {\bibinfo {volume} {05}},\ \bibinfo {pages} {005}
  (\bibinfo {year} {2012})},\ \Eprint {http://arxiv.org/abs/1203.2158}
  {arXiv:1203.2158 [hep-th]} \BibitemShut {NoStop}%
\bibitem [{\citenamefont {Litim}\ and\ \citenamefont
  {Satz}(2012)}]{Litim:2012vz}%
  \BibitemOpen
  \bibfield  {author} {\bibinfo {author} {\bibfnamefont {D.}~\bibnamefont
  {Litim}}\ and\ \bibinfo {author} {\bibfnamefont {A.}~\bibnamefont {Satz}},\
  }\href@noop {} {\  (\bibinfo {year} {2012})},\ \Eprint
  {http://arxiv.org/abs/1205.4218} {arXiv:1205.4218 [hep-th]} \BibitemShut
  {NoStop}%
\bibitem [{\citenamefont {Nink}\ and\ \citenamefont
  {Reuter}(2013)}]{Nink:2012kr}%
  \BibitemOpen
  \bibfield  {author} {\bibinfo {author} {\bibfnamefont {A.}~\bibnamefont
  {Nink}}\ and\ \bibinfo {author} {\bibfnamefont {M.}~\bibnamefont {Reuter}},\
  }\href {\doibase 10.1142/S0218271813300085} {\bibfield  {journal} {\bibinfo
  {journal} {Int.J.Mod.Phys.}\ }\textbf {\bibinfo {volume} {D22}},\ \bibinfo
  {pages} {138} (\bibinfo {year} {2013})},\ \Eprint
  {http://arxiv.org/abs/1212.4325} {arXiv:1212.4325 [hep-th]} \BibitemShut
  {NoStop}%
\bibitem [{\citenamefont {Rechenberger}\ and\ \citenamefont
  {Saueressig}(2013)}]{Rechenberger:2012dt}%
  \BibitemOpen
  \bibfield  {author} {\bibinfo {author} {\bibfnamefont {S.}~\bibnamefont
  {Rechenberger}}\ and\ \bibinfo {author} {\bibfnamefont {F.}~\bibnamefont
  {Saueressig}},\ }\href {\doibase 10.1007/JHEP03(2013)010} {\bibfield
  {journal} {\bibinfo  {journal} {JHEP}\ }\textbf {\bibinfo {volume} {03}},\
  \bibinfo {pages} {010} (\bibinfo {year} {2013})},\ \Eprint
  {http://arxiv.org/abs/1212.5114} {arXiv:1212.5114 [hep-th]} \BibitemShut
  {NoStop}%
\bibitem [{\citenamefont {Nink}(2015)}]{Nink:2014yya}%
  \BibitemOpen
  \bibfield  {author} {\bibinfo {author} {\bibfnamefont {A.}~\bibnamefont
  {Nink}},\ }\href {\doibase 10.1103/PhysRevD.91.044030} {\bibfield  {journal}
  {\bibinfo  {journal} {Phys.Rev.}\ }\textbf {\bibinfo {volume} {D91}},\
  \bibinfo {pages} {044030} (\bibinfo {year} {2015})},\ \Eprint
  {http://arxiv.org/abs/1410.7816} {arXiv:1410.7816 [hep-th]} \BibitemShut
  {NoStop}%
\bibitem [{\citenamefont {Falls}(2015{\natexlab{a}})}]{Falls:2015qga}%
  \BibitemOpen
  \bibfield  {author} {\bibinfo {author} {\bibfnamefont {K.}~\bibnamefont
  {Falls}},\ }\href {\doibase 10.1103/PhysRevD.92.124057} {\bibfield  {journal}
  {\bibinfo  {journal} {Phys. Rev.}\ }\textbf {\bibinfo {volume} {D92}},\
  \bibinfo {pages} {124057} (\bibinfo {year} {2015}{\natexlab{a}})},\ \Eprint
  {http://arxiv.org/abs/1501.05331} {arXiv:1501.05331 [hep-th]} \BibitemShut
  {NoStop}%
\bibitem [{\citenamefont {Falls}(2015{\natexlab{b}})}]{Falls:2015cta}%
  \BibitemOpen
  \bibfield  {author} {\bibinfo {author} {\bibfnamefont {K.}~\bibnamefont
  {Falls}},\ }\href@noop {} {\  (\bibinfo {year} {2015}{\natexlab{b}})},\
  \Eprint {http://arxiv.org/abs/1503.06233} {arXiv:1503.06233 [hep-th]}
  \BibitemShut {NoStop}%
\bibitem [{\citenamefont {Lauscher}\ and\ \citenamefont
  {Reuter}(2002{\natexlab{c}})}]{Lauscher:2002sq}%
  \BibitemOpen
  \bibfield  {author} {\bibinfo {author} {\bibfnamefont {O.}~\bibnamefont
  {Lauscher}}\ and\ \bibinfo {author} {\bibfnamefont {M.}~\bibnamefont
  {Reuter}},\ }\href {\doibase 10.1103/PhysRevD.66.025026} {\bibfield
  {journal} {\bibinfo  {journal} {Phys. Rev.}\ }\textbf {\bibinfo {volume}
  {D66}},\ \bibinfo {pages} {025026} (\bibinfo {year} {2002}{\natexlab{c}})},\
  \Eprint {http://arxiv.org/abs/hep-th/0205062} {arXiv:hep-th/0205062 [hep-th]}
  \BibitemShut {NoStop}%
\bibitem [{\citenamefont {Codello}\ and\ \citenamefont
  {Percacci}(2006)}]{Codello:2006in}%
  \BibitemOpen
  \bibfield  {author} {\bibinfo {author} {\bibfnamefont {A.}~\bibnamefont
  {Codello}}\ and\ \bibinfo {author} {\bibfnamefont {R.}~\bibnamefont
  {Percacci}},\ }\href {\doibase 10.1103/PhysRevLett.97.221301} {\bibfield
  {journal} {\bibinfo  {journal} {Phys. Rev. Lett.}\ }\textbf {\bibinfo
  {volume} {97}},\ \bibinfo {pages} {221301} (\bibinfo {year} {2006})},\
  \Eprint {http://arxiv.org/abs/hep-th/0607128} {arXiv:hep-th/0607128 [hep-th]}
  \BibitemShut {NoStop}%
\bibitem [{\citenamefont {Benedetti}\ \emph {et~al.}(2009)\citenamefont
  {Benedetti}, \citenamefont {Machado},\ and\ \citenamefont
  {Saueressig}}]{Benedetti:2009rx}%
  \BibitemOpen
  \bibfield  {author} {\bibinfo {author} {\bibfnamefont {D.}~\bibnamefont
  {Benedetti}}, \bibinfo {author} {\bibfnamefont {P.~F.}\ \bibnamefont
  {Machado}}, \ and\ \bibinfo {author} {\bibfnamefont {F.}~\bibnamefont
  {Saueressig}},\ }\href {\doibase 10.1142/S0217732309031521} {\bibfield
  {journal} {\bibinfo  {journal} {Mod. Phys. Lett.}\ }\textbf {\bibinfo
  {volume} {A24}},\ \bibinfo {pages} {2233} (\bibinfo {year} {2009})},\ \Eprint
  {http://arxiv.org/abs/0901.2984} {arXiv:0901.2984 [hep-th]} \BibitemShut
  {NoStop}%
\bibitem [{\citenamefont {Groh}\ \emph
  {et~al.}(2011{\natexlab{a}})\citenamefont {Groh}, \citenamefont
  {Rechenberger}, \citenamefont {Saueressig},\ and\ \citenamefont
  {Zanusso}}]{Groh:2011vn}%
  \BibitemOpen
  \bibfield  {author} {\bibinfo {author} {\bibfnamefont {K.}~\bibnamefont
  {Groh}}, \bibinfo {author} {\bibfnamefont {S.}~\bibnamefont {Rechenberger}},
  \bibinfo {author} {\bibfnamefont {F.}~\bibnamefont {Saueressig}}, \ and\
  \bibinfo {author} {\bibfnamefont {O.}~\bibnamefont {Zanusso}},\ }\href@noop
  {} {\bibfield  {journal} {\bibinfo  {journal} {PoS}\ }\textbf {\bibinfo
  {volume} {EPS-HEP2011}},\ \bibinfo {pages} {124} (\bibinfo {year}
  {2011}{\natexlab{a}})},\ \Eprint {http://arxiv.org/abs/1111.1743}
  {arXiv:1111.1743 [hep-th]} \BibitemShut {NoStop}%
\bibitem [{\citenamefont {Rechenberger}\ and\ \citenamefont
  {Saueressig}(2012)}]{Rechenberger:2012pm}%
  \BibitemOpen
  \bibfield  {author} {\bibinfo {author} {\bibfnamefont {S.}~\bibnamefont
  {Rechenberger}}\ and\ \bibinfo {author} {\bibfnamefont {F.}~\bibnamefont
  {Saueressig}},\ }\href {\doibase 10.1103/PhysRevD.86.024018} {\bibfield
  {journal} {\bibinfo  {journal} {Phys. Rev.}\ }\textbf {\bibinfo {volume}
  {D86}},\ \bibinfo {pages} {024018} (\bibinfo {year} {2012})},\ \Eprint
  {http://arxiv.org/abs/1206.0657} {arXiv:1206.0657 [hep-th]} \BibitemShut
  {NoStop}%
\bibitem [{\citenamefont {Ohta}\ and\ \citenamefont
  {Percacci}(2014)}]{Ohta:2013uca}%
  \BibitemOpen
  \bibfield  {author} {\bibinfo {author} {\bibfnamefont {N.}~\bibnamefont
  {Ohta}}\ and\ \bibinfo {author} {\bibfnamefont {R.}~\bibnamefont
  {Percacci}},\ }\href {\doibase 10.1088/0264-9381/31/1/015024} {\bibfield
  {journal} {\bibinfo  {journal} {Class. Quant. Grav.}\ }\textbf {\bibinfo
  {volume} {31}},\ \bibinfo {pages} {015024} (\bibinfo {year} {2014})},\
  \Eprint {http://arxiv.org/abs/1308.3398} {arXiv:1308.3398 [hep-th]}
  \BibitemShut {NoStop}%
\bibitem [{\citenamefont {Ohta}\ and\ \citenamefont
  {Percacci}(2016)}]{Ohta:2015zwa}%
  \BibitemOpen
  \bibfield  {author} {\bibinfo {author} {\bibfnamefont {N.}~\bibnamefont
  {Ohta}}\ and\ \bibinfo {author} {\bibfnamefont {R.}~\bibnamefont
  {Percacci}},\ }\href {\doibase 10.1088/0264-9381/33/3/035001} {\bibfield
  {journal} {\bibinfo  {journal} {Class. Quant. Grav.}\ }\textbf {\bibinfo
  {volume} {33}},\ \bibinfo {pages} {035001} (\bibinfo {year} {2016})},\
  \Eprint {http://arxiv.org/abs/1506.05526} {arXiv:1506.05526 [hep-th]}
  \BibitemShut {NoStop}%
\bibitem [{\citenamefont {Hamada}\ and\ \citenamefont
  {Yamada}(2017)}]{Hamada:2017rvn}%
  \BibitemOpen
  \bibfield  {author} {\bibinfo {author} {\bibfnamefont {Y.}~\bibnamefont
  {Hamada}}\ and\ \bibinfo {author} {\bibfnamefont {M.}~\bibnamefont
  {Yamada}},\ }\href@noop {} {\  (\bibinfo {year} {2017})},\ \Eprint
  {http://arxiv.org/abs/1703.09033} {arXiv:1703.09033 [hep-th]} \BibitemShut
  {NoStop}%
\bibitem [{\citenamefont {Christiansen}(2016)}]{Christiansen:2016sjn}%
  \BibitemOpen
  \bibfield  {author} {\bibinfo {author} {\bibfnamefont {N.}~\bibnamefont
  {Christiansen}},\ }\href@noop {} {\  (\bibinfo {year} {2016})},\ \Eprint
  {http://arxiv.org/abs/1612.06223} {arXiv:1612.06223 [hep-th]} \BibitemShut
  {NoStop}%
\bibitem [{\citenamefont {Machado}\ and\ \citenamefont
  {Saueressig}(2008)}]{Machado:2007ea}%
  \BibitemOpen
  \bibfield  {author} {\bibinfo {author} {\bibfnamefont {P.~F.}\ \bibnamefont
  {Machado}}\ and\ \bibinfo {author} {\bibfnamefont {F.}~\bibnamefont
  {Saueressig}},\ }\href {\doibase 10.1103/PhysRevD.77.124045} {\bibfield
  {journal} {\bibinfo  {journal} {Phys.Rev.}\ }\textbf {\bibinfo {volume}
  {D77}},\ \bibinfo {pages} {124045} (\bibinfo {year} {2008})},\ \Eprint
  {http://arxiv.org/abs/0712.0445} {arXiv:0712.0445 [hep-th]} \BibitemShut
  {NoStop}%
\bibitem [{\citenamefont {Codello}\ \emph {et~al.}(2008)\citenamefont
  {Codello}, \citenamefont {Percacci},\ and\ \citenamefont
  {Rahmede}}]{Codello:2007bd}%
  \BibitemOpen
  \bibfield  {author} {\bibinfo {author} {\bibfnamefont {A.}~\bibnamefont
  {Codello}}, \bibinfo {author} {\bibfnamefont {R.}~\bibnamefont {Percacci}}, \
  and\ \bibinfo {author} {\bibfnamefont {C.}~\bibnamefont {Rahmede}},\ }\href
  {\doibase 10.1142/S0217751X08038135} {\bibfield  {journal} {\bibinfo
  {journal} {Int.J.Mod.Phys.}\ }\textbf {\bibinfo {volume} {A23}},\ \bibinfo
  {pages} {143} (\bibinfo {year} {2008})},\ \Eprint
  {http://arxiv.org/abs/0705.1769} {arXiv:0705.1769 [hep-th]} \BibitemShut
  {NoStop}%
\bibitem [{\citenamefont {Bonanno}\ \emph {et~al.}(2011)\citenamefont
  {Bonanno}, \citenamefont {Contillo},\ and\ \citenamefont
  {Percacci}}]{Bonanno:2010bt}%
  \BibitemOpen
  \bibfield  {author} {\bibinfo {author} {\bibfnamefont {A.}~\bibnamefont
  {Bonanno}}, \bibinfo {author} {\bibfnamefont {A.}~\bibnamefont {Contillo}}, \
  and\ \bibinfo {author} {\bibfnamefont {R.}~\bibnamefont {Percacci}},\ }\href
  {\doibase 10.1088/0264-9381/28/14/145026} {\bibfield  {journal} {\bibinfo
  {journal} {Class. Quant. Grav.}\ }\textbf {\bibinfo {volume} {28}},\ \bibinfo
  {pages} {145026} (\bibinfo {year} {2011})},\ \Eprint
  {http://arxiv.org/abs/1006.0192} {arXiv:1006.0192 [gr-qc]} \BibitemShut
  {NoStop}%
\bibitem [{\citenamefont {Dietz}\ and\ \citenamefont
  {Morris}(2013{\natexlab{a}})}]{Dietz:2012ic}%
  \BibitemOpen
  \bibfield  {author} {\bibinfo {author} {\bibfnamefont {J.~A.}\ \bibnamefont
  {Dietz}}\ and\ \bibinfo {author} {\bibfnamefont {T.~R.}\ \bibnamefont
  {Morris}},\ }\href {\doibase 10.1007/JHEP01(2013)108} {\bibfield  {journal}
  {\bibinfo  {journal} {JHEP}\ }\textbf {\bibinfo {volume} {1301}},\ \bibinfo
  {pages} {108} (\bibinfo {year} {2013}{\natexlab{a}})},\ \Eprint
  {http://arxiv.org/abs/1211.0955} {arXiv:1211.0955 [hep-th]} \BibitemShut
  {NoStop}%
\bibitem [{\citenamefont {Demmel}\ \emph {et~al.}(2012)\citenamefont {Demmel},
  \citenamefont {Saueressig},\ and\ \citenamefont {Zanusso}}]{Demmel:2012ub}%
  \BibitemOpen
  \bibfield  {author} {\bibinfo {author} {\bibfnamefont {M.}~\bibnamefont
  {Demmel}}, \bibinfo {author} {\bibfnamefont {F.}~\bibnamefont {Saueressig}},
  \ and\ \bibinfo {author} {\bibfnamefont {O.}~\bibnamefont {Zanusso}},\ }\href
  {\doibase 10.1007/JHEP11(2012)131} {\bibfield  {journal} {\bibinfo  {journal}
  {JHEP}\ }\textbf {\bibinfo {volume} {11}},\ \bibinfo {pages} {131} (\bibinfo
  {year} {2012})},\ \Eprint {http://arxiv.org/abs/1208.2038} {arXiv:1208.2038
  [hep-th]} \BibitemShut {NoStop}%
\bibitem [{\citenamefont {Falls}\ \emph {et~al.}(2013)\citenamefont {Falls},
  \citenamefont {Litim}, \citenamefont {Nikolakopoulos},\ and\ \citenamefont
  {Rahmede}}]{Falls:2013bv}%
  \BibitemOpen
  \bibfield  {author} {\bibinfo {author} {\bibfnamefont {K.}~\bibnamefont
  {Falls}}, \bibinfo {author} {\bibfnamefont {D.}~\bibnamefont {Litim}},
  \bibinfo {author} {\bibfnamefont {K.}~\bibnamefont {Nikolakopoulos}}, \ and\
  \bibinfo {author} {\bibfnamefont {C.}~\bibnamefont {Rahmede}},\ }\href@noop
  {} {\  (\bibinfo {year} {2013})},\ \Eprint {http://arxiv.org/abs/1301.4191}
  {arXiv:1301.4191 [hep-th]} \BibitemShut {NoStop}%
\bibitem [{\citenamefont {Dietz}\ and\ \citenamefont
  {Morris}(2013{\natexlab{b}})}]{Dietz:2013sba}%
  \BibitemOpen
  \bibfield  {author} {\bibinfo {author} {\bibfnamefont {J.~A.}\ \bibnamefont
  {Dietz}}\ and\ \bibinfo {author} {\bibfnamefont {T.~R.}\ \bibnamefont
  {Morris}},\ }\href {\doibase 10.1007/JHEP07(2013)064} {\bibfield  {journal}
  {\bibinfo  {journal} {JHEP}\ }\textbf {\bibinfo {volume} {07}},\ \bibinfo
  {pages} {064} (\bibinfo {year} {2013}{\natexlab{b}})},\ \Eprint
  {http://arxiv.org/abs/1306.1223} {arXiv:1306.1223 [hep-th]} \BibitemShut
  {NoStop}%
\bibitem [{\citenamefont {Falls}\ \emph
  {et~al.}(2016{\natexlab{a}})\citenamefont {Falls}, \citenamefont {Litim},
  \citenamefont {Nikolakopoulos},\ and\ \citenamefont
  {Rahmede}}]{Falls:2014tra}%
  \BibitemOpen
  \bibfield  {author} {\bibinfo {author} {\bibfnamefont {K.}~\bibnamefont
  {Falls}}, \bibinfo {author} {\bibfnamefont {D.~F.}\ \bibnamefont {Litim}},
  \bibinfo {author} {\bibfnamefont {K.}~\bibnamefont {Nikolakopoulos}}, \ and\
  \bibinfo {author} {\bibfnamefont {C.}~\bibnamefont {Rahmede}},\ }\href
  {\doibase 10.1103/PhysRevD.93.104022} {\bibfield  {journal} {\bibinfo
  {journal} {Phys. Rev.}\ }\textbf {\bibinfo {volume} {D93}},\ \bibinfo {pages}
  {104022} (\bibinfo {year} {2016}{\natexlab{a}})},\ \Eprint
  {http://arxiv.org/abs/1410.4815} {arXiv:1410.4815 [hep-th]} \BibitemShut
  {NoStop}%
\bibitem [{\citenamefont {Demmel}\ \emph
  {et~al.}(2015{\natexlab{a}})\citenamefont {Demmel}, \citenamefont
  {Saueressig},\ and\ \citenamefont {Zanusso}}]{Demmel:2014hla}%
  \BibitemOpen
  \bibfield  {author} {\bibinfo {author} {\bibfnamefont {M.}~\bibnamefont
  {Demmel}}, \bibinfo {author} {\bibfnamefont {F.}~\bibnamefont {Saueressig}},
  \ and\ \bibinfo {author} {\bibfnamefont {O.}~\bibnamefont {Zanusso}},\ }\href
  {\doibase 10.1016/j.aop.2015.04.018} {\bibfield  {journal} {\bibinfo
  {journal} {Annals Phys.}\ }\textbf {\bibinfo {volume} {359}},\ \bibinfo
  {pages} {141} (\bibinfo {year} {2015}{\natexlab{a}})},\ \Eprint
  {http://arxiv.org/abs/1412.7207} {arXiv:1412.7207 [hep-th]} \BibitemShut
  {NoStop}%
\bibitem [{\citenamefont {Demmel}\ \emph {et~al.}(2014)\citenamefont {Demmel},
  \citenamefont {Saueressig},\ and\ \citenamefont {Zanusso}}]{Demmel:2014sga}%
  \BibitemOpen
  \bibfield  {author} {\bibinfo {author} {\bibfnamefont {M.}~\bibnamefont
  {Demmel}}, \bibinfo {author} {\bibfnamefont {F.}~\bibnamefont {Saueressig}},
  \ and\ \bibinfo {author} {\bibfnamefont {O.}~\bibnamefont {Zanusso}},\ }\href
  {\doibase 10.1007/JHEP06(2014)026} {\bibfield  {journal} {\bibinfo  {journal}
  {JHEP}\ }\textbf {\bibinfo {volume} {06}},\ \bibinfo {pages} {026} (\bibinfo
  {year} {2014})},\ \Eprint {http://arxiv.org/abs/1401.5495} {arXiv:1401.5495
  [hep-th]} \BibitemShut {NoStop}%
\bibitem [{\citenamefont {Eichhorn}(2015)}]{Eichhorn:2015bna}%
  \BibitemOpen
  \bibfield  {author} {\bibinfo {author} {\bibfnamefont {A.}~\bibnamefont
  {Eichhorn}},\ }\href {\doibase 10.1007/JHEP04(2015)096} {\bibfield  {journal}
  {\bibinfo  {journal} {JHEP}\ }\textbf {\bibinfo {volume} {1504}},\ \bibinfo
  {pages} {096} (\bibinfo {year} {2015})},\ \Eprint
  {http://arxiv.org/abs/1501.05848} {arXiv:1501.05848 [gr-qc]} \BibitemShut
  {NoStop}%
\bibitem [{\citenamefont {Demmel}\ \emph
  {et~al.}(2015{\natexlab{b}})\citenamefont {Demmel}, \citenamefont
  {Saueressig},\ and\ \citenamefont {Zanusso}}]{Demmel:2015oqa}%
  \BibitemOpen
  \bibfield  {author} {\bibinfo {author} {\bibfnamefont {M.}~\bibnamefont
  {Demmel}}, \bibinfo {author} {\bibfnamefont {F.}~\bibnamefont {Saueressig}},
  \ and\ \bibinfo {author} {\bibfnamefont {O.}~\bibnamefont {Zanusso}},\ }\href
  {\doibase 10.1007/JHEP08(2015)113} {\bibfield  {journal} {\bibinfo  {journal}
  {JHEP}\ }\textbf {\bibinfo {volume} {08}},\ \bibinfo {pages} {113} (\bibinfo
  {year} {2015}{\natexlab{b}})},\ \Eprint {http://arxiv.org/abs/1504.07656}
  {arXiv:1504.07656 [hep-th]} \BibitemShut {NoStop}%
\bibitem [{\citenamefont {Ohta}\ \emph {et~al.}(2015)\citenamefont {Ohta},
  \citenamefont {Percacci},\ and\ \citenamefont {Vacca}}]{Ohta:2015efa}%
  \BibitemOpen
  \bibfield  {author} {\bibinfo {author} {\bibfnamefont {N.}~\bibnamefont
  {Ohta}}, \bibinfo {author} {\bibfnamefont {R.}~\bibnamefont {Percacci}}, \
  and\ \bibinfo {author} {\bibfnamefont {G.~P.}\ \bibnamefont {Vacca}},\ }\href
  {\doibase 10.1103/PhysRevD.92.061501} {\bibfield  {journal} {\bibinfo
  {journal} {Phys. Rev.}\ }\textbf {\bibinfo {volume} {D92}},\ \bibinfo {pages}
  {061501} (\bibinfo {year} {2015})},\ \Eprint
  {http://arxiv.org/abs/1507.00968} {arXiv:1507.00968 [hep-th]} \BibitemShut
  {NoStop}%
\bibitem [{\citenamefont {Ohta}\ \emph
  {et~al.}(2016{\natexlab{a}})\citenamefont {Ohta}, \citenamefont {Percacci},\
  and\ \citenamefont {Vacca}}]{Ohta:2015fcu}%
  \BibitemOpen
  \bibfield  {author} {\bibinfo {author} {\bibfnamefont {N.}~\bibnamefont
  {Ohta}}, \bibinfo {author} {\bibfnamefont {R.}~\bibnamefont {Percacci}}, \
  and\ \bibinfo {author} {\bibfnamefont {G.~P.}\ \bibnamefont {Vacca}},\ }\href
  {\doibase 10.1140/epjc/s10052-016-3895-1} {\bibfield  {journal} {\bibinfo
  {journal} {Eur. Phys. J.}\ }\textbf {\bibinfo {volume} {C76}},\ \bibinfo
  {pages} {46} (\bibinfo {year} {2016}{\natexlab{a}})},\ \Eprint
  {http://arxiv.org/abs/1511.09393} {arXiv:1511.09393 [hep-th]} \BibitemShut
  {NoStop}%
\bibitem [{\citenamefont {Falls}\ \emph
  {et~al.}(2016{\natexlab{b}})\citenamefont {Falls}, \citenamefont {Litim},
  \citenamefont {Nikolakopoulos},\ and\ \citenamefont
  {Rahmede}}]{Falls:2016wsa}%
  \BibitemOpen
  \bibfield  {author} {\bibinfo {author} {\bibfnamefont {K.}~\bibnamefont
  {Falls}}, \bibinfo {author} {\bibfnamefont {D.~F.}\ \bibnamefont {Litim}},
  \bibinfo {author} {\bibfnamefont {K.}~\bibnamefont {Nikolakopoulos}}, \ and\
  \bibinfo {author} {\bibfnamefont {C.}~\bibnamefont {Rahmede}},\ }\href@noop
  {} {\  (\bibinfo {year} {2016}{\natexlab{b}})},\ \Eprint
  {http://arxiv.org/abs/1607.04962} {arXiv:1607.04962 [gr-qc]} \BibitemShut
  {NoStop}%
\bibitem [{\citenamefont {Falls}\ and\ \citenamefont
  {Ohta}(2016)}]{Falls:2016msz}%
  \BibitemOpen
  \bibfield  {author} {\bibinfo {author} {\bibfnamefont {K.}~\bibnamefont
  {Falls}}\ and\ \bibinfo {author} {\bibfnamefont {N.}~\bibnamefont {Ohta}},\
  }\href {\doibase 10.1103/PhysRevD.94.084005} {\bibfield  {journal} {\bibinfo
  {journal} {Phys. Rev.}\ }\textbf {\bibinfo {volume} {D94}},\ \bibinfo {pages}
  {084005} (\bibinfo {year} {2016})},\ \Eprint
  {http://arxiv.org/abs/1607.08460} {arXiv:1607.08460 [hep-th]} \BibitemShut
  {NoStop}%
\bibitem [{\citenamefont {Morris}(2016)}]{Morris:2016spn}%
  \BibitemOpen
  \bibfield  {author} {\bibinfo {author} {\bibfnamefont {T.~R.}\ \bibnamefont
  {Morris}},\ }\href {\doibase 10.1007/JHEP11(2016)160} {\bibfield  {journal}
  {\bibinfo  {journal} {JHEP}\ }\textbf {\bibinfo {volume} {11}},\ \bibinfo
  {pages} {160} (\bibinfo {year} {2016})},\ \Eprint
  {http://arxiv.org/abs/1610.03081} {arXiv:1610.03081 [hep-th]} \BibitemShut
  {NoStop}%
\bibitem [{\citenamefont {Gonzalez-Martin}\ \emph {et~al.}(2017)\citenamefont
  {Gonzalez-Martin}, \citenamefont {Morris},\ and\ \citenamefont
  {Slade}}]{Gonzalez-Martin:2017gza}%
  \BibitemOpen
  \bibfield  {author} {\bibinfo {author} {\bibfnamefont {S.}~\bibnamefont
  {Gonzalez-Martin}}, \bibinfo {author} {\bibfnamefont {T.~R.}\ \bibnamefont
  {Morris}}, \ and\ \bibinfo {author} {\bibfnamefont {Z.~H.}\ \bibnamefont
  {Slade}},\ }\href {\doibase 10.1103/PhysRevD.95.106010} {\bibfield  {journal}
  {\bibinfo  {journal} {Phys. Rev.}\ }\textbf {\bibinfo {volume} {D95}},\
  \bibinfo {pages} {106010} (\bibinfo {year} {2017})},\ \Eprint
  {http://arxiv.org/abs/1704.08873} {arXiv:1704.08873 [hep-th]} \BibitemShut
  {NoStop}%
\bibitem [{\citenamefont {Gies}\ \emph {et~al.}(2016)\citenamefont {Gies},
  \citenamefont {Knorr}, \citenamefont {Lippoldt},\ and\ \citenamefont
  {Saueressig}}]{Gies:2016con}%
  \BibitemOpen
  \bibfield  {author} {\bibinfo {author} {\bibfnamefont {H.}~\bibnamefont
  {Gies}}, \bibinfo {author} {\bibfnamefont {B.}~\bibnamefont {Knorr}},
  \bibinfo {author} {\bibfnamefont {S.}~\bibnamefont {Lippoldt}}, \ and\
  \bibinfo {author} {\bibfnamefont {F.}~\bibnamefont {Saueressig}},\ }\href
  {\doibase 10.1103/PhysRevLett.116.211302} {\bibfield  {journal} {\bibinfo
  {journal} {Phys. Rev. Lett.}\ }\textbf {\bibinfo {volume} {116}},\ \bibinfo
  {pages} {211302} (\bibinfo {year} {2016})},\ \Eprint
  {http://arxiv.org/abs/1601.01800} {arXiv:1601.01800 [hep-th]} \BibitemShut
  {NoStop}%
\bibitem [{\citenamefont {Biemans}\ \emph
  {et~al.}(2017{\natexlab{a}})\citenamefont {Biemans}, \citenamefont
  {Platania},\ and\ \citenamefont {Saueressig}}]{Biemans:2016rvp}%
  \BibitemOpen
  \bibfield  {author} {\bibinfo {author} {\bibfnamefont {J.}~\bibnamefont
  {Biemans}}, \bibinfo {author} {\bibfnamefont {A.}~\bibnamefont {Platania}}, \
  and\ \bibinfo {author} {\bibfnamefont {F.}~\bibnamefont {Saueressig}},\
  }\href {\doibase 10.1103/PhysRevD.95.086013} {\bibfield  {journal} {\bibinfo
  {journal} {Phys. Rev.}\ }\textbf {\bibinfo {volume} {D95}},\ \bibinfo {pages}
  {086013} (\bibinfo {year} {2017}{\natexlab{a}})},\ \Eprint
  {http://arxiv.org/abs/1609.04813} {arXiv:1609.04813 [hep-th]} \BibitemShut
  {NoStop}%
\bibitem [{\citenamefont {Biemans}\ \emph
  {et~al.}(2017{\natexlab{b}})\citenamefont {Biemans}, \citenamefont
  {Platania},\ and\ \citenamefont {Saueressig}}]{Biemans:2017zca}%
  \BibitemOpen
  \bibfield  {author} {\bibinfo {author} {\bibfnamefont {J.}~\bibnamefont
  {Biemans}}, \bibinfo {author} {\bibfnamefont {A.}~\bibnamefont {Platania}}, \
  and\ \bibinfo {author} {\bibfnamefont {F.}~\bibnamefont {Saueressig}},\
  }\href {\doibase 10.1007/JHEP05(2017)093} {\bibfield  {journal} {\bibinfo
  {journal} {JHEP}\ }\textbf {\bibinfo {volume} {05}},\ \bibinfo {pages} {093}
  (\bibinfo {year} {2017}{\natexlab{b}})},\ \Eprint
  {http://arxiv.org/abs/1702.06539} {arXiv:1702.06539 [hep-th]} \BibitemShut
  {NoStop}%
\bibitem [{\citenamefont {Houthoff}\ \emph {et~al.}(2017)\citenamefont
  {Houthoff}, \citenamefont {Kurov},\ and\ \citenamefont
  {Saueressig}}]{Houthoff:2017oam}%
  \BibitemOpen
  \bibfield  {author} {\bibinfo {author} {\bibfnamefont {W.~B.}\ \bibnamefont
  {Houthoff}}, \bibinfo {author} {\bibfnamefont {A.}~\bibnamefont {Kurov}}, \
  and\ \bibinfo {author} {\bibfnamefont {F.}~\bibnamefont {Saueressig}},\
  }\href@noop {} {\  (\bibinfo {year} {2017})},\ \Eprint
  {http://arxiv.org/abs/1705.01848} {arXiv:1705.01848 [hep-th]} \BibitemShut
  {NoStop}%
\bibitem [{\citenamefont {Pagani}\ and\ \citenamefont
  {Percacci}(2015)}]{Pagani:2015ema}%
  \BibitemOpen
  \bibfield  {author} {\bibinfo {author} {\bibfnamefont {C.}~\bibnamefont
  {Pagani}}\ and\ \bibinfo {author} {\bibfnamefont {R.}~\bibnamefont
  {Percacci}},\ }\href {\doibase 10.1088/0264-9381/32/19/195019} {\bibfield
  {journal} {\bibinfo  {journal} {Class. Quant. Grav.}\ }\textbf {\bibinfo
  {volume} {32}},\ \bibinfo {pages} {195019} (\bibinfo {year} {2015})},\
  \Eprint {http://arxiv.org/abs/1506.02882} {arXiv:1506.02882 [gr-qc]}
  \BibitemShut {NoStop}%
\bibitem [{\citenamefont {Nink}\ and\ \citenamefont
  {Reuter}(2016)}]{Nink:2015lmq}%
  \BibitemOpen
  \bibfield  {author} {\bibinfo {author} {\bibfnamefont {A.}~\bibnamefont
  {Nink}}\ and\ \bibinfo {author} {\bibfnamefont {M.}~\bibnamefont {Reuter}},\
  }\href {\doibase 10.1007/JHEP02(2016)167} {\bibfield  {journal} {\bibinfo
  {journal} {JHEP}\ }\textbf {\bibinfo {volume} {02}},\ \bibinfo {pages} {167}
  (\bibinfo {year} {2016})},\ \Eprint {http://arxiv.org/abs/1512.06805}
  {arXiv:1512.06805 [hep-th]} \BibitemShut {NoStop}%
\bibitem [{\citenamefont {Percacci}\ and\ \citenamefont
  {Perini}(2003{\natexlab{a}})}]{Percacci:2002ie}%
  \BibitemOpen
  \bibfield  {author} {\bibinfo {author} {\bibfnamefont {R.}~\bibnamefont
  {Percacci}}\ and\ \bibinfo {author} {\bibfnamefont {D.}~\bibnamefont
  {Perini}},\ }\href {\doibase 10.1103/PhysRevD.67.081503} {\bibfield
  {journal} {\bibinfo  {journal} {Phys. Rev.}\ }\textbf {\bibinfo {volume}
  {D67}},\ \bibinfo {pages} {081503} (\bibinfo {year} {2003}{\natexlab{a}})},\
  \Eprint {http://arxiv.org/abs/hep-th/0207033} {arXiv:hep-th/0207033 [hep-th]}
  \BibitemShut {NoStop}%
\bibitem [{\citenamefont {Percacci}\ and\ \citenamefont
  {Perini}(2003{\natexlab{b}})}]{Percacci:2003jz}%
  \BibitemOpen
  \bibfield  {author} {\bibinfo {author} {\bibfnamefont {R.}~\bibnamefont
  {Percacci}}\ and\ \bibinfo {author} {\bibfnamefont {D.}~\bibnamefont
  {Perini}},\ }\href {\doibase 10.1103/PhysRevD.68.044018} {\bibfield
  {journal} {\bibinfo  {journal} {Phys. Rev.}\ }\textbf {\bibinfo {volume}
  {D68}},\ \bibinfo {pages} {044018} (\bibinfo {year} {2003}{\natexlab{b}})},\
  \Eprint {http://arxiv.org/abs/hep-th/0304222} {arXiv:hep-th/0304222 [hep-th]}
  \BibitemShut {NoStop}%
\bibitem [{\citenamefont {Daum}\ \emph {et~al.}(2010)\citenamefont {Daum},
  \citenamefont {Harst},\ and\ \citenamefont {Reuter}}]{Daum:2009dn}%
  \BibitemOpen
  \bibfield  {author} {\bibinfo {author} {\bibfnamefont {J.-E.}\ \bibnamefont
  {Daum}}, \bibinfo {author} {\bibfnamefont {U.}~\bibnamefont {Harst}}, \ and\
  \bibinfo {author} {\bibfnamefont {M.}~\bibnamefont {Reuter}},\ }\href
  {\doibase 10.1007/JHEP01(2010)084} {\bibfield  {journal} {\bibinfo  {journal}
  {JHEP}\ }\textbf {\bibinfo {volume} {01}},\ \bibinfo {pages} {084} (\bibinfo
  {year} {2010})},\ \Eprint {http://arxiv.org/abs/0910.4938} {arXiv:0910.4938
  [hep-th]} \BibitemShut {NoStop}%
\bibitem [{\citenamefont {Vacca}\ and\ \citenamefont
  {Zanusso}(2010)}]{Vacca:2010mj}%
  \BibitemOpen
  \bibfield  {author} {\bibinfo {author} {\bibfnamefont {G.~P.}\ \bibnamefont
  {Vacca}}\ and\ \bibinfo {author} {\bibfnamefont {O.}~\bibnamefont
  {Zanusso}},\ }\href {\doibase 10.1103/PhysRevLett.105.231601} {\bibfield
  {journal} {\bibinfo  {journal} {Phys. Rev. Lett.}\ }\textbf {\bibinfo
  {volume} {105}},\ \bibinfo {pages} {231601} (\bibinfo {year} {2010})},\
  \Eprint {http://arxiv.org/abs/1009.1735} {arXiv:1009.1735 [hep-th]}
  \BibitemShut {NoStop}%
\bibitem [{\citenamefont {Folkerts}\ \emph {et~al.}(2012)\citenamefont
  {Folkerts}, \citenamefont {Litim},\ and\ \citenamefont
  {Pawlowski}}]{Folkerts:2011jz}%
  \BibitemOpen
  \bibfield  {author} {\bibinfo {author} {\bibfnamefont {S.}~\bibnamefont
  {Folkerts}}, \bibinfo {author} {\bibfnamefont {D.~F.}\ \bibnamefont {Litim}},
  \ and\ \bibinfo {author} {\bibfnamefont {J.~M.}\ \bibnamefont {Pawlowski}},\
  }\href {\doibase 10.1016/j.physletb.2012.02.002} {\bibfield  {journal}
  {\bibinfo  {journal} {Phys.Lett.}\ }\textbf {\bibinfo {volume} {B709}},\
  \bibinfo {pages} {234} (\bibinfo {year} {2012})},\ \Eprint
  {http://arxiv.org/abs/1101.5552} {arXiv:1101.5552 [hep-th]} \BibitemShut
  {NoStop}%
\bibitem [{\citenamefont {Eichhorn}\ and\ \citenamefont
  {Gies}(2011)}]{Eichhorn:2011pc}%
  \BibitemOpen
  \bibfield  {author} {\bibinfo {author} {\bibfnamefont {A.}~\bibnamefont
  {Eichhorn}}\ and\ \bibinfo {author} {\bibfnamefont {H.}~\bibnamefont
  {Gies}},\ }\href {\doibase 10.1088/1367-2630/13/12/125012} {\bibfield
  {journal} {\bibinfo  {journal} {New J.Phys.}\ }\textbf {\bibinfo {volume}
  {13}},\ \bibinfo {pages} {125012} (\bibinfo {year} {2011})},\ \Eprint
  {http://arxiv.org/abs/1104.5366} {arXiv:1104.5366 [hep-th]} \BibitemShut
  {NoStop}%
\bibitem [{\citenamefont {Harst}\ and\ \citenamefont
  {Reuter}(2011)}]{Harst:2011zx}%
  \BibitemOpen
  \bibfield  {author} {\bibinfo {author} {\bibfnamefont {U.}~\bibnamefont
  {Harst}}\ and\ \bibinfo {author} {\bibfnamefont {M.}~\bibnamefont {Reuter}},\
  }\href {\doibase 10.1007/JHEP05(2011)119} {\bibfield  {journal} {\bibinfo
  {journal} {JHEP}\ }\textbf {\bibinfo {volume} {1105}},\ \bibinfo {pages}
  {119} (\bibinfo {year} {2011})},\ \Eprint {http://arxiv.org/abs/1101.6007}
  {arXiv:1101.6007 [hep-th]} \BibitemShut {NoStop}%
\bibitem [{\citenamefont {Eichhorn}(2012)}]{Eichhorn:2012va}%
  \BibitemOpen
  \bibfield  {author} {\bibinfo {author} {\bibfnamefont {A.}~\bibnamefont
  {Eichhorn}},\ }\href {\doibase 10.1103/PhysRevD.86.105021} {\bibfield
  {journal} {\bibinfo  {journal} {Phys. Rev.}\ }\textbf {\bibinfo {volume}
  {D86}},\ \bibinfo {pages} {105021} (\bibinfo {year} {2012})},\ \Eprint
  {http://arxiv.org/abs/1204.0965} {arXiv:1204.0965 [gr-qc]} \BibitemShut
  {NoStop}%
\bibitem [{\citenamefont {Dobrich}\ and\ \citenamefont
  {Eichhorn}(2012)}]{Dobrich:2012nv}%
  \BibitemOpen
  \bibfield  {author} {\bibinfo {author} {\bibfnamefont {B.}~\bibnamefont
  {Dobrich}}\ and\ \bibinfo {author} {\bibfnamefont {A.}~\bibnamefont
  {Eichhorn}},\ }\href {\doibase 10.1007/JHEP06(2012)156} {\bibfield  {journal}
  {\bibinfo  {journal} {JHEP}\ }\textbf {\bibinfo {volume} {06}},\ \bibinfo
  {pages} {156} (\bibinfo {year} {2012})},\ \Eprint
  {http://arxiv.org/abs/1203.6366} {arXiv:1203.6366 [gr-qc]} \BibitemShut
  {NoStop}%
\bibitem [{\citenamefont {Dona}\ and\ \citenamefont
  {Percacci}(2013)}]{Dona:2012am}%
  \BibitemOpen
  \bibfield  {author} {\bibinfo {author} {\bibfnamefont {P.}~\bibnamefont
  {Dona}}\ and\ \bibinfo {author} {\bibfnamefont {R.}~\bibnamefont
  {Percacci}},\ }\href {\doibase 10.1103/PhysRevD.87.045002} {\bibfield
  {journal} {\bibinfo  {journal} {Phys. Rev.}\ }\textbf {\bibinfo {volume}
  {D87}},\ \bibinfo {pages} {045002} (\bibinfo {year} {2013})},\ \Eprint
  {http://arxiv.org/abs/1209.3649} {arXiv:1209.3649 [hep-th]} \BibitemShut
  {NoStop}%
\bibitem [{\citenamefont {Don\`a}\ \emph {et~al.}(2014)\citenamefont {Don\`a},
  \citenamefont {Eichhorn},\ and\ \citenamefont {Percacci}}]{Dona:2013qba}%
  \BibitemOpen
  \bibfield  {author} {\bibinfo {author} {\bibfnamefont {P.}~\bibnamefont
  {Don\`a}}, \bibinfo {author} {\bibfnamefont {A.}~\bibnamefont {Eichhorn}}, \
  and\ \bibinfo {author} {\bibfnamefont {R.}~\bibnamefont {Percacci}},\ }\href
  {\doibase 10.1103/PhysRevD.89.084035} {\bibfield  {journal} {\bibinfo
  {journal} {Phys.Rev.}\ }\textbf {\bibinfo {volume} {D89}},\ \bibinfo {pages}
  {084035} (\bibinfo {year} {2014})},\ \Eprint {http://arxiv.org/abs/1311.2898}
  {arXiv:1311.2898 [hep-th]} \BibitemShut {NoStop}%
\bibitem [{\citenamefont {Percacci}\ and\ \citenamefont
  {Vacca}(2015)}]{Percacci:2015wwa}%
  \BibitemOpen
  \bibfield  {author} {\bibinfo {author} {\bibfnamefont {R.}~\bibnamefont
  {Percacci}}\ and\ \bibinfo {author} {\bibfnamefont {G.~P.}\ \bibnamefont
  {Vacca}},\ }\href {\doibase 10.1140/epjc/s10052-015-3410-0} {\bibfield
  {journal} {\bibinfo  {journal} {Eur. Phys. J.}\ }\textbf {\bibinfo {volume}
  {C75}},\ \bibinfo {pages} {188} (\bibinfo {year} {2015})},\ \Eprint
  {http://arxiv.org/abs/1501.00888} {arXiv:1501.00888 [hep-th]} \BibitemShut
  {NoStop}%
\bibitem [{\citenamefont {Borchardt}\ and\ \citenamefont
  {Knorr}(2015)}]{Borchardt:2015rxa}%
  \BibitemOpen
  \bibfield  {author} {\bibinfo {author} {\bibfnamefont {J.}~\bibnamefont
  {Borchardt}}\ and\ \bibinfo {author} {\bibfnamefont {B.}~\bibnamefont
  {Knorr}},\ }\href {\doibase 10.1103/PhysRevD.93.089904,
  10.1103/PhysRevD.91.105011} {\bibfield  {journal} {\bibinfo  {journal} {Phys.
  Rev.}\ }\textbf {\bibinfo {volume} {D91}},\ \bibinfo {pages} {105011}
  (\bibinfo {year} {2015})},\ \bibinfo {note} {[Erratum: Phys.
  Rev.D93,no.8,089904(2016)]},\ \Eprint {http://arxiv.org/abs/1502.07511}
  {arXiv:1502.07511 [hep-th]} \BibitemShut {NoStop}%
\bibitem [{\citenamefont {Meibohm}\ \emph {et~al.}(2016)\citenamefont
  {Meibohm}, \citenamefont {Pawlowski},\ and\ \citenamefont
  {Reichert}}]{Meibohm:2015twa}%
  \BibitemOpen
  \bibfield  {author} {\bibinfo {author} {\bibfnamefont {J.}~\bibnamefont
  {Meibohm}}, \bibinfo {author} {\bibfnamefont {J.~M.}\ \bibnamefont
  {Pawlowski}}, \ and\ \bibinfo {author} {\bibfnamefont {M.}~\bibnamefont
  {Reichert}},\ }\href {\doibase 10.1103/PhysRevD.93.084035} {\bibfield
  {journal} {\bibinfo  {journal} {Phys. Rev.}\ }\textbf {\bibinfo {volume}
  {D93}},\ \bibinfo {pages} {084035} (\bibinfo {year} {2016})},\ \Eprint
  {http://arxiv.org/abs/1510.07018} {arXiv:1510.07018 [hep-th]} \BibitemShut
  {NoStop}%
\bibitem [{\citenamefont {Donà}\ \emph {et~al.}(2016)\citenamefont {Donà},
  \citenamefont {Eichhorn}, \citenamefont {Labus},\ and\ \citenamefont
  {Percacci}}]{Dona:2015tnf}%
  \BibitemOpen
  \bibfield  {author} {\bibinfo {author} {\bibfnamefont {P.}~\bibnamefont
  {Donà}}, \bibinfo {author} {\bibfnamefont {A.}~\bibnamefont {Eichhorn}},
  \bibinfo {author} {\bibfnamefont {P.}~\bibnamefont {Labus}}, \ and\ \bibinfo
  {author} {\bibfnamefont {R.}~\bibnamefont {Percacci}},\ }\href {\doibase
  10.1103/PhysRevD.93.129904, 10.1103/PhysRevD.93.044049} {\bibfield  {journal}
  {\bibinfo  {journal} {Phys. Rev.}\ }\textbf {\bibinfo {volume} {D93}},\
  \bibinfo {pages} {044049} (\bibinfo {year} {2016})},\ \bibinfo {note}
  {[Erratum: Phys. Rev.D93,no.12,129904(2016)]},\ \Eprint
  {http://arxiv.org/abs/1512.01589} {arXiv:1512.01589 [gr-qc]} \BibitemShut
  {NoStop}%
\bibitem [{\citenamefont {Labus}\ \emph
  {et~al.}(2016{\natexlab{a}})\citenamefont {Labus}, \citenamefont {Percacci},\
  and\ \citenamefont {Vacca}}]{Labus:2015ska}%
  \BibitemOpen
  \bibfield  {author} {\bibinfo {author} {\bibfnamefont {P.}~\bibnamefont
  {Labus}}, \bibinfo {author} {\bibfnamefont {R.}~\bibnamefont {Percacci}}, \
  and\ \bibinfo {author} {\bibfnamefont {G.~P.}\ \bibnamefont {Vacca}},\ }\href
  {\doibase 10.1016/j.physletb.2015.12.022} {\bibfield  {journal} {\bibinfo
  {journal} {Phys. Lett.}\ }\textbf {\bibinfo {volume} {B753}},\ \bibinfo
  {pages} {274} (\bibinfo {year} {2016}{\natexlab{a}})},\ \Eprint
  {http://arxiv.org/abs/1505.05393} {arXiv:1505.05393 [hep-th]} \BibitemShut
  {NoStop}%
\bibitem [{\citenamefont {Meibohm}\ and\ \citenamefont
  {Pawlowski}(2016)}]{Meibohm:2016mkp}%
  \BibitemOpen
  \bibfield  {author} {\bibinfo {author} {\bibfnamefont {J.}~\bibnamefont
  {Meibohm}}\ and\ \bibinfo {author} {\bibfnamefont {J.~M.}\ \bibnamefont
  {Pawlowski}},\ }\href {\doibase 10.1140/epjc/s10052-016-4132-7} {\bibfield
  {journal} {\bibinfo  {journal} {Eur. Phys. J.}\ }\textbf {\bibinfo {volume}
  {C76}},\ \bibinfo {pages} {285} (\bibinfo {year} {2016})},\ \Eprint
  {http://arxiv.org/abs/1601.04597} {arXiv:1601.04597 [hep-th]} \BibitemShut
  {NoStop}%
\bibitem [{\citenamefont {Eichhorn}\ \emph {et~al.}(2016)\citenamefont
  {Eichhorn}, \citenamefont {Held},\ and\ \citenamefont
  {Pawlowski}}]{Eichhorn:2016esv}%
  \BibitemOpen
  \bibfield  {author} {\bibinfo {author} {\bibfnamefont {A.}~\bibnamefont
  {Eichhorn}}, \bibinfo {author} {\bibfnamefont {A.}~\bibnamefont {Held}}, \
  and\ \bibinfo {author} {\bibfnamefont {J.~M.}\ \bibnamefont {Pawlowski}},\
  }\href {\doibase 10.1103/PhysRevD.94.104027} {\bibfield  {journal} {\bibinfo
  {journal} {Phys. Rev.}\ }\textbf {\bibinfo {volume} {D94}},\ \bibinfo {pages}
  {104027} (\bibinfo {year} {2016})},\ \Eprint
  {http://arxiv.org/abs/1604.02041} {arXiv:1604.02041 [hep-th]} \BibitemShut
  {NoStop}%
\bibitem [{\citenamefont {Eichhorn}\ and\ \citenamefont
  {Lippoldt}(2017)}]{Eichhorn:2016vvy}%
  \BibitemOpen
  \bibfield  {author} {\bibinfo {author} {\bibfnamefont {A.}~\bibnamefont
  {Eichhorn}}\ and\ \bibinfo {author} {\bibfnamefont {S.}~\bibnamefont
  {Lippoldt}},\ }\href {\doibase 10.1016/j.physletb.2017.01.064} {\bibfield
  {journal} {\bibinfo  {journal} {Phys. Lett.}\ }\textbf {\bibinfo {volume}
  {B767}},\ \bibinfo {pages} {142} (\bibinfo {year} {2017})},\ \Eprint
  {http://arxiv.org/abs/1611.05878} {arXiv:1611.05878 [gr-qc]} \BibitemShut
  {NoStop}%
\bibitem [{\citenamefont {Christiansen}\ and\ \citenamefont
  {Eichhorn}(2017)}]{Christiansen:2017gtg}%
  \BibitemOpen
  \bibfield  {author} {\bibinfo {author} {\bibfnamefont {N.}~\bibnamefont
  {Christiansen}}\ and\ \bibinfo {author} {\bibfnamefont {A.}~\bibnamefont
  {Eichhorn}},\ }\href {\doibase 10.1016/j.physletb.2017.04.047} {\bibfield
  {journal} {\bibinfo  {journal} {Phys. Lett.}\ }\textbf {\bibinfo {volume}
  {B770}},\ \bibinfo {pages} {154} (\bibinfo {year} {2017})},\ \Eprint
  {http://arxiv.org/abs/1702.07724} {arXiv:1702.07724 [hep-th]} \BibitemShut
  {NoStop}%
\bibitem [{\citenamefont {Christiansen}\ \emph {et~al.}(2017)\citenamefont
  {Christiansen}, \citenamefont {Eichhorn},\ and\ \citenamefont
  {Held}}]{Christiansen:2017qca}%
  \BibitemOpen
  \bibfield  {author} {\bibinfo {author} {\bibfnamefont {N.}~\bibnamefont
  {Christiansen}}, \bibinfo {author} {\bibfnamefont {A.}~\bibnamefont
  {Eichhorn}}, \ and\ \bibinfo {author} {\bibfnamefont {A.}~\bibnamefont
  {Held}},\ }\href@noop {} {\  (\bibinfo {year} {2017})},\ \Eprint
  {http://arxiv.org/abs/1705.01858} {arXiv:1705.01858 [hep-th]} \BibitemShut
  {NoStop}%
\bibitem [{\citenamefont {Eichhorn}\ and\ \citenamefont
  {Held}(2017)}]{Eichhorn:2017eht}%
  \BibitemOpen
  \bibfield  {author} {\bibinfo {author} {\bibfnamefont {A.}~\bibnamefont
  {Eichhorn}}\ and\ \bibinfo {author} {\bibfnamefont {A.}~\bibnamefont
  {Held}},\ }\href@noop {} {\  (\bibinfo {year} {2017})},\ \Eprint
  {http://arxiv.org/abs/1705.02342} {arXiv:1705.02342 [gr-qc]} \BibitemShut
  {NoStop}%
\bibitem [{\citenamefont {Bonanno}\ and\ \citenamefont
  {Reuter}(2000)}]{Bonanno:2000ep}%
  \BibitemOpen
  \bibfield  {author} {\bibinfo {author} {\bibfnamefont {A.}~\bibnamefont
  {Bonanno}}\ and\ \bibinfo {author} {\bibfnamefont {M.}~\bibnamefont
  {Reuter}},\ }\href {\doibase 10.1103/PhysRevD.62.043008} {\bibfield
  {journal} {\bibinfo  {journal} {Phys. Rev.}\ }\textbf {\bibinfo {volume}
  {D62}},\ \bibinfo {pages} {043008} (\bibinfo {year} {2000})},\ \Eprint
  {http://arxiv.org/abs/hep-th/0002196} {arXiv:hep-th/0002196 [hep-th]}
  \BibitemShut {NoStop}%
\bibitem [{\citenamefont {Falls}\ \emph {et~al.}(2012)\citenamefont {Falls},
  \citenamefont {Litim},\ and\ \citenamefont {Raghuraman}}]{Falls:2010he}%
  \BibitemOpen
  \bibfield  {author} {\bibinfo {author} {\bibfnamefont {K.}~\bibnamefont
  {Falls}}, \bibinfo {author} {\bibfnamefont {D.~F.}\ \bibnamefont {Litim}}, \
  and\ \bibinfo {author} {\bibfnamefont {A.}~\bibnamefont {Raghuraman}},\
  }\href {\doibase 10.1142/S0217751X12500194} {\bibfield  {journal} {\bibinfo
  {journal} {Int. J. Mod. Phys.}\ }\textbf {\bibinfo {volume} {A27}},\ \bibinfo
  {pages} {1250019} (\bibinfo {year} {2012})},\ \Eprint
  {http://arxiv.org/abs/1002.0260} {arXiv:1002.0260 [hep-th]} \BibitemShut
  {NoStop}%
\bibitem [{\citenamefont {Falls}\ and\ \citenamefont
  {Litim}(2014)}]{Falls:2012nd}%
  \BibitemOpen
  \bibfield  {author} {\bibinfo {author} {\bibfnamefont {K.}~\bibnamefont
  {Falls}}\ and\ \bibinfo {author} {\bibfnamefont {D.~F.}\ \bibnamefont
  {Litim}},\ }\href {\doibase 10.1103/PhysRevD.89.084002} {\bibfield  {journal}
  {\bibinfo  {journal} {Phys. Rev.}\ }\textbf {\bibinfo {volume} {D89}},\
  \bibinfo {pages} {084002} (\bibinfo {year} {2014})},\ \Eprint
  {http://arxiv.org/abs/1212.1821} {arXiv:1212.1821 [gr-qc]} \BibitemShut
  {NoStop}%
\bibitem [{\citenamefont {Becker}\ and\ \citenamefont
  {Reuter}(2012)}]{Becker:2012js}%
  \BibitemOpen
  \bibfield  {author} {\bibinfo {author} {\bibfnamefont {D.}~\bibnamefont
  {Becker}}\ and\ \bibinfo {author} {\bibfnamefont {M.}~\bibnamefont
  {Reuter}},\ }\href {\doibase 10.1007/JHEP07(2012)172} {\bibfield  {journal}
  {\bibinfo  {journal} {JHEP}\ }\textbf {\bibinfo {volume} {07}},\ \bibinfo
  {pages} {172} (\bibinfo {year} {2012})},\ \Eprint
  {http://arxiv.org/abs/1205.3583} {arXiv:1205.3583 [hep-th]} \BibitemShut
  {NoStop}%
\bibitem [{\citenamefont {Koch}\ and\ \citenamefont
  {Saueressig}(2014{\natexlab{a}})}]{Koch:2013owa}%
  \BibitemOpen
  \bibfield  {author} {\bibinfo {author} {\bibfnamefont {B.}~\bibnamefont
  {Koch}}\ and\ \bibinfo {author} {\bibfnamefont {F.}~\bibnamefont
  {Saueressig}},\ }\href {\doibase 10.1088/0264-9381/31/1/015006} {\bibfield
  {journal} {\bibinfo  {journal} {Class. Quant. Grav.}\ }\textbf {\bibinfo
  {volume} {31}},\ \bibinfo {pages} {015006} (\bibinfo {year}
  {2014}{\natexlab{a}})},\ \Eprint {http://arxiv.org/abs/1306.1546}
  {arXiv:1306.1546 [hep-th]} \BibitemShut {NoStop}%
\bibitem [{\citenamefont {Koch}\ and\ \citenamefont
  {Saueressig}(2014{\natexlab{b}})}]{Koch:2014cqa}%
  \BibitemOpen
  \bibfield  {author} {\bibinfo {author} {\bibfnamefont {B.}~\bibnamefont
  {Koch}}\ and\ \bibinfo {author} {\bibfnamefont {F.}~\bibnamefont
  {Saueressig}},\ }\href {\doibase 10.1142/S0217751X14300117} {\bibfield
  {journal} {\bibinfo  {journal} {Int. J. Mod. Phys.}\ }\textbf {\bibinfo
  {volume} {A29}},\ \bibinfo {pages} {1430011} (\bibinfo {year}
  {2014}{\natexlab{b}})},\ \Eprint {http://arxiv.org/abs/1401.4452}
  {arXiv:1401.4452 [hep-th]} \BibitemShut {NoStop}%
\bibitem [{\citenamefont {Bonanno}\ and\ \citenamefont
  {Reuter}(2002)}]{Bonanno:2001xi}%
  \BibitemOpen
  \bibfield  {author} {\bibinfo {author} {\bibfnamefont {A.}~\bibnamefont
  {Bonanno}}\ and\ \bibinfo {author} {\bibfnamefont {M.}~\bibnamefont
  {Reuter}},\ }\href {\doibase 10.1103/PhysRevD.65.043508} {\bibfield
  {journal} {\bibinfo  {journal} {Phys. Rev.}\ }\textbf {\bibinfo {volume}
  {D65}},\ \bibinfo {pages} {043508} (\bibinfo {year} {2002})},\ \Eprint
  {http://arxiv.org/abs/hep-th/0106133} {arXiv:hep-th/0106133 [hep-th]}
  \BibitemShut {NoStop}%
\bibitem [{\citenamefont {Reuter}\ and\ \citenamefont
  {Saueressig}(2005)}]{Reuter:2005kb}%
  \BibitemOpen
  \bibfield  {author} {\bibinfo {author} {\bibfnamefont {M.}~\bibnamefont
  {Reuter}}\ and\ \bibinfo {author} {\bibfnamefont {F.}~\bibnamefont
  {Saueressig}},\ }\href {\doibase 10.1088/1475-7516/2005/09/012} {\bibfield
  {journal} {\bibinfo  {journal} {JCAP}\ }\textbf {\bibinfo {volume} {0509}},\
  \bibinfo {pages} {012} (\bibinfo {year} {2005})},\ \Eprint
  {http://arxiv.org/abs/hep-th/0507167} {arXiv:hep-th/0507167 [hep-th]}
  \BibitemShut {NoStop}%
\bibitem [{\citenamefont {Bonanno}\ and\ \citenamefont
  {Reuter}(2007)}]{Bonanno:2007wg}%
  \BibitemOpen
  \bibfield  {author} {\bibinfo {author} {\bibfnamefont {A.}~\bibnamefont
  {Bonanno}}\ and\ \bibinfo {author} {\bibfnamefont {M.}~\bibnamefont
  {Reuter}},\ }\href {\doibase 10.1088/1475-7516/2007/08/024} {\bibfield
  {journal} {\bibinfo  {journal} {JCAP}\ }\textbf {\bibinfo {volume} {0708}},\
  \bibinfo {pages} {024} (\bibinfo {year} {2007})},\ \Eprint
  {http://arxiv.org/abs/0706.0174} {arXiv:0706.0174 [hep-th]} \BibitemShut
  {NoStop}%
\bibitem [{\citenamefont {Hindmarsh}\ \emph {et~al.}(2011)\citenamefont
  {Hindmarsh}, \citenamefont {Litim},\ and\ \citenamefont
  {Rahmede}}]{Hindmarsh:2011hx}%
  \BibitemOpen
  \bibfield  {author} {\bibinfo {author} {\bibfnamefont {M.}~\bibnamefont
  {Hindmarsh}}, \bibinfo {author} {\bibfnamefont {D.}~\bibnamefont {Litim}}, \
  and\ \bibinfo {author} {\bibfnamefont {C.}~\bibnamefont {Rahmede}},\ }\href
  {\doibase 10.1088/1475-7516/2011/07/019} {\bibfield  {journal} {\bibinfo
  {journal} {JCAP}\ }\textbf {\bibinfo {volume} {1107}},\ \bibinfo {pages}
  {019} (\bibinfo {year} {2011})},\ \Eprint {http://arxiv.org/abs/1101.5401}
  {arXiv:1101.5401 [gr-qc]} \BibitemShut {NoStop}%
\bibitem [{\citenamefont {Henz}\ \emph {et~al.}(2013)\citenamefont {Henz},
  \citenamefont {Pawlowski}, \citenamefont {Rodigast},\ and\ \citenamefont
  {Wetterich}}]{Henz:2013oxa}%
  \BibitemOpen
  \bibfield  {author} {\bibinfo {author} {\bibfnamefont {T.}~\bibnamefont
  {Henz}}, \bibinfo {author} {\bibfnamefont {J.~M.}\ \bibnamefont {Pawlowski}},
  \bibinfo {author} {\bibfnamefont {A.}~\bibnamefont {Rodigast}}, \ and\
  \bibinfo {author} {\bibfnamefont {C.}~\bibnamefont {Wetterich}},\ }\href
  {\doibase 10.1016/j.physletb.2013.10.015} {\bibfield  {journal} {\bibinfo
  {journal} {Phys.Lett.}\ }\textbf {\bibinfo {volume} {B727}},\ \bibinfo
  {pages} {298} (\bibinfo {year} {2013})},\ \Eprint
  {http://arxiv.org/abs/1304.7743} {arXiv:1304.7743 [hep-th]} \BibitemShut
  {NoStop}%
\bibitem [{\citenamefont {Saltas}(2016)}]{Saltas:2015vsc}%
  \BibitemOpen
  \bibfield  {author} {\bibinfo {author} {\bibfnamefont {I.~D.}\ \bibnamefont
  {Saltas}},\ }\href {\doibase 10.1088/1475-7516/2016/02/048} {\bibfield
  {journal} {\bibinfo  {journal} {JCAP}\ }\textbf {\bibinfo {volume} {1602}},\
  \bibinfo {pages} {048} (\bibinfo {year} {2016})},\ \Eprint
  {http://arxiv.org/abs/1512.06134} {arXiv:1512.06134 [hep-th]} \BibitemShut
  {NoStop}%
\bibitem [{\citenamefont {Bonanno}\ and\ \citenamefont
  {Platania}(2015)}]{Bonanno:2015fga}%
  \BibitemOpen
  \bibfield  {author} {\bibinfo {author} {\bibfnamefont {A.}~\bibnamefont
  {Bonanno}}\ and\ \bibinfo {author} {\bibfnamefont {A.}~\bibnamefont
  {Platania}},\ }\href {\doibase 10.1016/j.physletb.2015.10.005} {\bibfield
  {journal} {\bibinfo  {journal} {Phys. Lett.}\ }\textbf {\bibinfo {volume}
  {B750}},\ \bibinfo {pages} {638} (\bibinfo {year} {2015})},\ \Eprint
  {http://arxiv.org/abs/1507.03375} {arXiv:1507.03375 [gr-qc]} \BibitemShut
  {NoStop}%
\bibitem [{\citenamefont {Henz}\ \emph {et~al.}(2017)\citenamefont {Henz},
  \citenamefont {Pawlowski},\ and\ \citenamefont {Wetterich}}]{Henz:2016aoh}%
  \BibitemOpen
  \bibfield  {author} {\bibinfo {author} {\bibfnamefont {T.}~\bibnamefont
  {Henz}}, \bibinfo {author} {\bibfnamefont {J.~M.}\ \bibnamefont {Pawlowski}},
  \ and\ \bibinfo {author} {\bibfnamefont {C.}~\bibnamefont {Wetterich}},\
  }\href {\doibase 10.1016/j.physletb.2017.01.057} {\bibfield  {journal}
  {\bibinfo  {journal} {Phys. Lett.}\ }\textbf {\bibinfo {volume} {B769}},\
  \bibinfo {pages} {105} (\bibinfo {year} {2017})},\ \Eprint
  {http://arxiv.org/abs/1605.01858} {arXiv:1605.01858 [hep-th]} \BibitemShut
  {NoStop}%
\bibitem [{\citenamefont {Bonanno}\ \emph {et~al.}(2017)\citenamefont
  {Bonanno}, \citenamefont {Koch},\ and\ \citenamefont
  {Platania}}]{Bonanno:2016dyv}%
  \BibitemOpen
  \bibfield  {author} {\bibinfo {author} {\bibfnamefont {A.}~\bibnamefont
  {Bonanno}}, \bibinfo {author} {\bibfnamefont {B.}~\bibnamefont {Koch}}, \
  and\ \bibinfo {author} {\bibfnamefont {A.}~\bibnamefont {Platania}},\ }\href
  {\doibase 10.1088/1361-6382/aa6788} {\bibfield  {journal} {\bibinfo
  {journal} {Class. Quant. Grav.}\ }\textbf {\bibinfo {volume} {34}},\ \bibinfo
  {pages} {095012} (\bibinfo {year} {2017})},\ \Eprint
  {http://arxiv.org/abs/1610.05299} {arXiv:1610.05299 [gr-qc]} \BibitemShut
  {NoStop}%
\bibitem [{\citenamefont {Wetterich}(2017{\natexlab{a}})}]{Wetterich:2017ixo}%
  \BibitemOpen
  \bibfield  {author} {\bibinfo {author} {\bibfnamefont {C.}~\bibnamefont
  {Wetterich}},\ }\href@noop {} {\  (\bibinfo {year} {2017}{\natexlab{a}})},\
  \Eprint {http://arxiv.org/abs/1704.08040} {arXiv:1704.08040 [gr-qc]}
  \BibitemShut {NoStop}%
\bibitem [{\citenamefont {Bonanno}\ and\ \citenamefont
  {Saueressig}(2017)}]{Bonanno:2017pkg}%
  \BibitemOpen
  \bibfield  {author} {\bibinfo {author} {\bibfnamefont {A.}~\bibnamefont
  {Bonanno}}\ and\ \bibinfo {author} {\bibfnamefont {F.}~\bibnamefont
  {Saueressig}},\ }\href {\doibase 10.1016/j.crhy.2017.02.002} {\bibfield
  {journal} {\bibinfo  {journal} {Comptes Rendus Physique}\ }\textbf {\bibinfo
  {volume} {18}},\ \bibinfo {pages} {254} (\bibinfo {year} {2017})},\ \Eprint
  {http://arxiv.org/abs/1702.04137} {arXiv:1702.04137 [hep-th]} \BibitemShut
  {NoStop}%
\bibitem [{\citenamefont {Alkofer}\ \emph {et~al.}(2016)\citenamefont
  {Alkofer}, \citenamefont {D'Odorico}, \citenamefont {Saueressig},\ and\
  \citenamefont {Versteegen}}]{Alkofer:2016utc}%
  \BibitemOpen
  \bibfield  {author} {\bibinfo {author} {\bibfnamefont {N.}~\bibnamefont
  {Alkofer}}, \bibinfo {author} {\bibfnamefont {G.}~\bibnamefont {D'Odorico}},
  \bibinfo {author} {\bibfnamefont {F.}~\bibnamefont {Saueressig}}, \ and\
  \bibinfo {author} {\bibfnamefont {F.}~\bibnamefont {Versteegen}},\ }\href
  {\doibase 10.1103/PhysRevD.94.104055} {\bibfield  {journal} {\bibinfo
  {journal} {Phys. Rev.}\ }\textbf {\bibinfo {volume} {D94}},\ \bibinfo {pages}
  {104055} (\bibinfo {year} {2016})},\ \Eprint
  {http://arxiv.org/abs/1605.08015} {arXiv:1605.08015 [gr-qc]} \BibitemShut
  {NoStop}%
\bibitem [{\citenamefont {Becker}\ and\ \citenamefont
  {Reuter}(2015)}]{Becker:2014pea}%
  \BibitemOpen
  \bibfield  {author} {\bibinfo {author} {\bibfnamefont {D.}~\bibnamefont
  {Becker}}\ and\ \bibinfo {author} {\bibfnamefont {M.}~\bibnamefont
  {Reuter}},\ }\href {\doibase 10.1007/JHEP03(2015)065} {\bibfield  {journal}
  {\bibinfo  {journal} {JHEP}\ }\textbf {\bibinfo {volume} {03}},\ \bibinfo
  {pages} {065} (\bibinfo {year} {2015})},\ \Eprint
  {http://arxiv.org/abs/1412.0468} {arXiv:1412.0468 [hep-th]} \BibitemShut
  {NoStop}%
\bibitem [{\citenamefont {Shaposhnikov}\ and\ \citenamefont
  {Wetterich}(2010)}]{Shaposhnikov:2009pv}%
  \BibitemOpen
  \bibfield  {author} {\bibinfo {author} {\bibfnamefont {M.}~\bibnamefont
  {Shaposhnikov}}\ and\ \bibinfo {author} {\bibfnamefont {C.}~\bibnamefont
  {Wetterich}},\ }\href {\doibase 10.1016/j.physletb.2009.12.022} {\bibfield
  {journal} {\bibinfo  {journal} {Phys. Lett.}\ }\textbf {\bibinfo {volume}
  {B683}},\ \bibinfo {pages} {196} (\bibinfo {year} {2010})},\ \Eprint
  {http://arxiv.org/abs/0912.0208} {arXiv:0912.0208 [hep-th]} \BibitemShut
  {NoStop}%
\bibitem [{\citenamefont {Chatrchyan}\ \emph {et~al.}(2012)\citenamefont
  {Chatrchyan} \emph {et~al.}}]{Chatrchyan:2012xdj}%
  \BibitemOpen
  \bibfield  {author} {\bibinfo {author} {\bibfnamefont {S.}~\bibnamefont
  {Chatrchyan}} \emph {et~al.} (\bibinfo {collaboration} {CMS}),\ }\href
  {\doibase 10.1016/j.physletb.2012.08.021} {\bibfield  {journal} {\bibinfo
  {journal} {Phys. Lett.}\ }\textbf {\bibinfo {volume} {B716}},\ \bibinfo
  {pages} {30} (\bibinfo {year} {2012})},\ \Eprint
  {http://arxiv.org/abs/1207.7235} {arXiv:1207.7235 [hep-ex]} \BibitemShut
  {NoStop}%
\bibitem [{\citenamefont {Aad}\ \emph {et~al.}(2012)\citenamefont {Aad} \emph
  {et~al.}}]{Aad:2012tfa}%
  \BibitemOpen
  \bibfield  {author} {\bibinfo {author} {\bibfnamefont {G.}~\bibnamefont
  {Aad}} \emph {et~al.} (\bibinfo {collaboration} {ATLAS}),\ }\href {\doibase
  10.1016/j.physletb.2012.08.020} {\bibfield  {journal} {\bibinfo  {journal}
  {Phys. Lett.}\ }\textbf {\bibinfo {volume} {B716}},\ \bibinfo {pages} {1}
  (\bibinfo {year} {2012})},\ \Eprint {http://arxiv.org/abs/1207.7214}
  {arXiv:1207.7214 [hep-ex]} \BibitemShut {NoStop}%
\bibitem [{\citenamefont {Litim}\ and\ \citenamefont
  {Pawlowski}(2002)}]{Litim:2002ce}%
  \BibitemOpen
  \bibfield  {author} {\bibinfo {author} {\bibfnamefont {D.~F.}\ \bibnamefont
  {Litim}}\ and\ \bibinfo {author} {\bibfnamefont {J.~M.}\ \bibnamefont
  {Pawlowski}},\ }\href {\doibase 10.1088/1126-6708/2002/09/049} {\bibfield
  {journal} {\bibinfo  {journal} {JHEP}\ }\textbf {\bibinfo {volume} {09}},\
  \bibinfo {pages} {049} (\bibinfo {year} {2002})},\ \Eprint
  {http://arxiv.org/abs/hep-th/0203005} {arXiv:hep-th/0203005 [hep-th]}
  \BibitemShut {NoStop}%
\bibitem [{\citenamefont {Bridle}\ \emph {et~al.}(2014)\citenamefont {Bridle},
  \citenamefont {Dietz},\ and\ \citenamefont {Morris}}]{Bridle:2013sra}%
  \BibitemOpen
  \bibfield  {author} {\bibinfo {author} {\bibfnamefont {I.~H.}\ \bibnamefont
  {Bridle}}, \bibinfo {author} {\bibfnamefont {J.~A.}\ \bibnamefont {Dietz}}, \
  and\ \bibinfo {author} {\bibfnamefont {T.~R.}\ \bibnamefont {Morris}},\
  }\href {\doibase 10.1007/JHEP03(2014)093} {\bibfield  {journal} {\bibinfo
  {journal} {JHEP}\ }\textbf {\bibinfo {volume} {1403}},\ \bibinfo {pages}
  {093} (\bibinfo {year} {2014})},\ \Eprint {http://arxiv.org/abs/1312.2846}
  {arXiv:1312.2846 [hep-th]} \BibitemShut {NoStop}%
\bibitem [{\citenamefont {Manrique}\ \emph
  {et~al.}(2011{\natexlab{b}})\citenamefont {Manrique}, \citenamefont
  {Reuter},\ and\ \citenamefont {Saueressig}}]{Manrique:2010mq}%
  \BibitemOpen
  \bibfield  {author} {\bibinfo {author} {\bibfnamefont {E.}~\bibnamefont
  {Manrique}}, \bibinfo {author} {\bibfnamefont {M.}~\bibnamefont {Reuter}}, \
  and\ \bibinfo {author} {\bibfnamefont {F.}~\bibnamefont {Saueressig}},\
  }\href {\doibase 10.1016/j.aop.2010.11.003} {\bibfield  {journal} {\bibinfo
  {journal} {Annals Phys.}\ }\textbf {\bibinfo {volume} {326}},\ \bibinfo
  {pages} {440} (\bibinfo {year} {2011}{\natexlab{b}})},\ \Eprint
  {http://arxiv.org/abs/1003.5129} {arXiv:1003.5129 [hep-th]} \BibitemShut
  {NoStop}%
\bibitem [{\citenamefont {Manrique}\ and\ \citenamefont
  {Reuter}(2010)}]{Manrique:2009uh}%
  \BibitemOpen
  \bibfield  {author} {\bibinfo {author} {\bibfnamefont {E.}~\bibnamefont
  {Manrique}}\ and\ \bibinfo {author} {\bibfnamefont {M.}~\bibnamefont
  {Reuter}},\ }\href {\doibase 10.1016/j.aop.2009.11.009} {\bibfield  {journal}
  {\bibinfo  {journal} {Annals Phys.}\ }\textbf {\bibinfo {volume} {325}},\
  \bibinfo {pages} {785} (\bibinfo {year} {2010})},\ \Eprint
  {http://arxiv.org/abs/0907.2617} {arXiv:0907.2617 [gr-qc]} \BibitemShut
  {NoStop}%
\bibitem [{\citenamefont {Manrique}\ \emph
  {et~al.}(2011{\natexlab{c}})\citenamefont {Manrique}, \citenamefont
  {Reuter},\ and\ \citenamefont {Saueressig}}]{Manrique:2010am}%
  \BibitemOpen
  \bibfield  {author} {\bibinfo {author} {\bibfnamefont {E.}~\bibnamefont
  {Manrique}}, \bibinfo {author} {\bibfnamefont {M.}~\bibnamefont {Reuter}}, \
  and\ \bibinfo {author} {\bibfnamefont {F.}~\bibnamefont {Saueressig}},\
  }\href {\doibase 10.1016/j.aop.2010.11.006} {\bibfield  {journal} {\bibinfo
  {journal} {Annals Phys.}\ }\textbf {\bibinfo {volume} {326}},\ \bibinfo
  {pages} {463} (\bibinfo {year} {2011}{\natexlab{c}})},\ \Eprint
  {http://arxiv.org/abs/1006.0099} {arXiv:1006.0099 [hep-th]} \BibitemShut
  {NoStop}%
\bibitem [{\citenamefont {Christiansen}\ \emph {et~al.}(2014)\citenamefont
  {Christiansen}, \citenamefont {Litim}, \citenamefont {Pawlowski},\ and\
  \citenamefont {Rodigast}}]{Christiansen:2012rx}%
  \BibitemOpen
  \bibfield  {author} {\bibinfo {author} {\bibfnamefont {N.}~\bibnamefont
  {Christiansen}}, \bibinfo {author} {\bibfnamefont {D.~F.}\ \bibnamefont
  {Litim}}, \bibinfo {author} {\bibfnamefont {J.~M.}\ \bibnamefont
  {Pawlowski}}, \ and\ \bibinfo {author} {\bibfnamefont {A.}~\bibnamefont
  {Rodigast}},\ }\href {\doibase 10.1016/j.physletb.2013.11.025} {\bibfield
  {journal} {\bibinfo  {journal} {Phys.Lett.}\ }\textbf {\bibinfo {volume}
  {B728}},\ \bibinfo {pages} {114} (\bibinfo {year} {2014})},\ \Eprint
  {http://arxiv.org/abs/1209.4038} {arXiv:1209.4038 [hep-th]} \BibitemShut
  {NoStop}%
\bibitem [{\citenamefont {Christiansen}\ \emph {et~al.}(2016)\citenamefont
  {Christiansen}, \citenamefont {Knorr}, \citenamefont {Pawlowski},\ and\
  \citenamefont {Rodigast}}]{Christiansen:2014raa}%
  \BibitemOpen
  \bibfield  {author} {\bibinfo {author} {\bibfnamefont {N.}~\bibnamefont
  {Christiansen}}, \bibinfo {author} {\bibfnamefont {B.}~\bibnamefont {Knorr}},
  \bibinfo {author} {\bibfnamefont {J.~M.}\ \bibnamefont {Pawlowski}}, \ and\
  \bibinfo {author} {\bibfnamefont {A.}~\bibnamefont {Rodigast}},\ }\href
  {\doibase 10.1103/PhysRevD.93.044036} {\bibfield  {journal} {\bibinfo
  {journal} {Phys. Rev.}\ }\textbf {\bibinfo {volume} {D93}},\ \bibinfo {pages}
  {044036} (\bibinfo {year} {2016})},\ \Eprint {http://arxiv.org/abs/1403.1232}
  {arXiv:1403.1232 [hep-th]} \BibitemShut {NoStop}%
\bibitem [{\citenamefont {Becker}\ and\ \citenamefont
  {Reuter}(2014{\natexlab{a}})}]{Becker:2014qya}%
  \BibitemOpen
  \bibfield  {author} {\bibinfo {author} {\bibfnamefont {D.}~\bibnamefont
  {Becker}}\ and\ \bibinfo {author} {\bibfnamefont {M.}~\bibnamefont
  {Reuter}},\ }\href {\doibase 10.1016/j.aop.2014.07.023} {\bibfield  {journal}
  {\bibinfo  {journal} {Annals Phys.}\ }\textbf {\bibinfo {volume} {350}},\
  \bibinfo {pages} {225} (\bibinfo {year} {2014}{\natexlab{a}})},\ \Eprint
  {http://arxiv.org/abs/1404.4537} {arXiv:1404.4537 [hep-th]} \BibitemShut
  {NoStop}%
\bibitem [{\citenamefont {Becker}\ and\ \citenamefont
  {Reuter}(2014{\natexlab{b}})}]{Becker:2014jua}%
  \BibitemOpen
  \bibfield  {author} {\bibinfo {author} {\bibfnamefont {D.}~\bibnamefont
  {Becker}}\ and\ \bibinfo {author} {\bibfnamefont {M.}~\bibnamefont
  {Reuter}},\ }\href {\doibase 10.1007/JHEP12(2014)025} {\bibfield  {journal}
  {\bibinfo  {journal} {JHEP}\ }\textbf {\bibinfo {volume} {12}},\ \bibinfo
  {pages} {025} (\bibinfo {year} {2014}{\natexlab{b}})},\ \Eprint
  {http://arxiv.org/abs/1407.5848} {arXiv:1407.5848 [hep-th]} \BibitemShut
  {NoStop}%
\bibitem [{\citenamefont {Christiansen}\ \emph {et~al.}(2015)\citenamefont
  {Christiansen}, \citenamefont {Knorr}, \citenamefont {Meibohm}, \citenamefont
  {Pawlowski},\ and\ \citenamefont {Reichert}}]{Christiansen:2015rva}%
  \BibitemOpen
  \bibfield  {author} {\bibinfo {author} {\bibfnamefont {N.}~\bibnamefont
  {Christiansen}}, \bibinfo {author} {\bibfnamefont {B.}~\bibnamefont {Knorr}},
  \bibinfo {author} {\bibfnamefont {J.}~\bibnamefont {Meibohm}}, \bibinfo
  {author} {\bibfnamefont {J.~M.}\ \bibnamefont {Pawlowski}}, \ and\ \bibinfo
  {author} {\bibfnamefont {M.}~\bibnamefont {Reichert}},\ }\href {\doibase
  10.1103/PhysRevD.92.121501} {\bibfield  {journal} {\bibinfo  {journal} {Phys.
  Rev.}\ }\textbf {\bibinfo {volume} {D92}},\ \bibinfo {pages} {121501}
  (\bibinfo {year} {2015})},\ \Eprint {http://arxiv.org/abs/1506.07016}
  {arXiv:1506.07016 [hep-th]} \BibitemShut {NoStop}%
\bibitem [{\citenamefont {Denz}\ \emph {et~al.}(2016)\citenamefont {Denz},
  \citenamefont {Pawlowski},\ and\ \citenamefont {Reichert}}]{Denz:2016qks}%
  \BibitemOpen
  \bibfield  {author} {\bibinfo {author} {\bibfnamefont {T.}~\bibnamefont
  {Denz}}, \bibinfo {author} {\bibfnamefont {J.~M.}\ \bibnamefont {Pawlowski}},
  \ and\ \bibinfo {author} {\bibfnamefont {M.}~\bibnamefont {Reichert}},\
  }\href@noop {} {\  (\bibinfo {year} {2016})},\ \Eprint
  {http://arxiv.org/abs/1612.07315} {arXiv:1612.07315 [hep-th]} \BibitemShut
  {NoStop}%
\bibitem [{\citenamefont {Pawlowski}(2003)}]{Pawlowski:2003sk}%
  \BibitemOpen
  \bibfield  {author} {\bibinfo {author} {\bibfnamefont {J.~M.}\ \bibnamefont
  {Pawlowski}},\ }\href@noop {} {\  (\bibinfo {year} {2003})},\ \Eprint
  {http://arxiv.org/abs/hep-th/0310018} {arXiv:hep-th/0310018 [hep-th]}
  \BibitemShut {NoStop}%
\bibitem [{\citenamefont {Pawlowski}(2007)}]{Pawlowski:2005xe}%
  \BibitemOpen
  \bibfield  {author} {\bibinfo {author} {\bibfnamefont {J.~M.}\ \bibnamefont
  {Pawlowski}},\ }\href {\doibase 10.1016/j.aop.2007.01.007} {\bibfield
  {journal} {\bibinfo  {journal} {Annals Phys.}\ }\textbf {\bibinfo {volume}
  {322}},\ \bibinfo {pages} {2831} (\bibinfo {year} {2007})},\ \Eprint
  {http://arxiv.org/abs/hep-th/0512261} {arXiv:hep-th/0512261 [hep-th]}
  \BibitemShut {NoStop}%
\bibitem [{\citenamefont {Donkin}\ and\ \citenamefont
  {Pawlowski}(2012)}]{Donkin:2012ud}%
  \BibitemOpen
  \bibfield  {author} {\bibinfo {author} {\bibfnamefont {I.}~\bibnamefont
  {Donkin}}\ and\ \bibinfo {author} {\bibfnamefont {J.~M.}\ \bibnamefont
  {Pawlowski}},\ }\href@noop {} {\  (\bibinfo {year} {2012})},\ \Eprint
  {http://arxiv.org/abs/1203.4207} {arXiv:1203.4207 [hep-th]} \BibitemShut
  {NoStop}%
\bibitem [{\citenamefont {Dietz}\ and\ \citenamefont
  {Morris}(2015)}]{Dietz:2015owa}%
  \BibitemOpen
  \bibfield  {author} {\bibinfo {author} {\bibfnamefont {J.~A.}\ \bibnamefont
  {Dietz}}\ and\ \bibinfo {author} {\bibfnamefont {T.~R.}\ \bibnamefont
  {Morris}},\ }\href {\doibase 10.1007/JHEP04(2015)118} {\bibfield  {journal}
  {\bibinfo  {journal} {JHEP}\ }\textbf {\bibinfo {volume} {04}},\ \bibinfo
  {pages} {118} (\bibinfo {year} {2015})},\ \Eprint
  {http://arxiv.org/abs/1502.07396} {arXiv:1502.07396 [hep-th]} \BibitemShut
  {NoStop}%
\bibitem [{\citenamefont {Labus}\ \emph
  {et~al.}(2016{\natexlab{b}})\citenamefont {Labus}, \citenamefont {Morris},\
  and\ \citenamefont {Slade}}]{Labus:2016lkh}%
  \BibitemOpen
  \bibfield  {author} {\bibinfo {author} {\bibfnamefont {P.}~\bibnamefont
  {Labus}}, \bibinfo {author} {\bibfnamefont {T.~R.}\ \bibnamefont {Morris}}, \
  and\ \bibinfo {author} {\bibfnamefont {Z.~H.}\ \bibnamefont {Slade}},\ }\href
  {\doibase 10.1103/PhysRevD.94.024007} {\bibfield  {journal} {\bibinfo
  {journal} {Phys. Rev.}\ }\textbf {\bibinfo {volume} {D94}},\ \bibinfo {pages}
  {024007} (\bibinfo {year} {2016}{\natexlab{b}})},\ \Eprint
  {http://arxiv.org/abs/1603.04772} {arXiv:1603.04772 [hep-th]} \BibitemShut
  {NoStop}%
\bibitem [{\citenamefont {Morris}\ and\ \citenamefont
  {Preston}(2016)}]{Morris:2016nda}%
  \BibitemOpen
  \bibfield  {author} {\bibinfo {author} {\bibfnamefont {T.~R.}\ \bibnamefont
  {Morris}}\ and\ \bibinfo {author} {\bibfnamefont {A.~W.~H.}\ \bibnamefont
  {Preston}},\ }\href {\doibase 10.1007/JHEP06(2016)012} {\bibfield  {journal}
  {\bibinfo  {journal} {JHEP}\ }\textbf {\bibinfo {volume} {06}},\ \bibinfo
  {pages} {012} (\bibinfo {year} {2016})},\ \Eprint
  {http://arxiv.org/abs/1602.08993} {arXiv:1602.08993 [hep-th]} \BibitemShut
  {NoStop}%
\bibitem [{\citenamefont {Percacci}\ and\ \citenamefont
  {Vacca}(2017)}]{Percacci:2016arh}%
  \BibitemOpen
  \bibfield  {author} {\bibinfo {author} {\bibfnamefont {R.}~\bibnamefont
  {Percacci}}\ and\ \bibinfo {author} {\bibfnamefont {G.~P.}\ \bibnamefont
  {Vacca}},\ }\href {\doibase 10.1140/epjc/s10052-017-4619-x} {\bibfield
  {journal} {\bibinfo  {journal} {Eur. Phys. J.}\ }\textbf {\bibinfo {volume}
  {C77}},\ \bibinfo {pages} {52} (\bibinfo {year} {2017})},\ \Eprint
  {http://arxiv.org/abs/1611.07005} {arXiv:1611.07005 [hep-th]} \BibitemShut
  {NoStop}%
\bibitem [{\citenamefont {Ohta}(2017)}]{Ohta:2017dsq}%
  \BibitemOpen
  \bibfield  {author} {\bibinfo {author} {\bibfnamefont {N.}~\bibnamefont
  {Ohta}},\ }\href {\doibase 10.1093/ptep/ptx020} {\bibfield  {journal}
  {\bibinfo  {journal} {PTEP}\ }\textbf {\bibinfo {volume} {2017}},\ \bibinfo
  {pages} {033E02} (\bibinfo {year} {2017})},\ \Eprint
  {http://arxiv.org/abs/1701.01506} {arXiv:1701.01506 [hep-th]} \BibitemShut
  {NoStop}%
\bibitem [{\citenamefont {Wetterich}(2016)}]{Wetterich:2016ewc}%
  \BibitemOpen
  \bibfield  {author} {\bibinfo {author} {\bibfnamefont {C.}~\bibnamefont
  {Wetterich}},\ }\href@noop {} {\  (\bibinfo {year} {2016})},\ \Eprint
  {http://arxiv.org/abs/1607.02989} {arXiv:1607.02989 [hep-th]} \BibitemShut
  {NoStop}%
\bibitem [{\citenamefont {Wetterich}(2017{\natexlab{b}})}]{Wetterich:2016qee}%
  \BibitemOpen
  \bibfield  {author} {\bibinfo {author} {\bibfnamefont {C.}~\bibnamefont
  {Wetterich}},\ }\href {\doibase 10.1016/j.nuclphysb.2016.12.008} {\bibfield
  {journal} {\bibinfo  {journal} {Nucl. Phys.}\ }\textbf {\bibinfo {volume}
  {B915}},\ \bibinfo {pages} {135} (\bibinfo {year} {2017}{\natexlab{b}})},\
  \Eprint {http://arxiv.org/abs/1608.01515} {arXiv:1608.01515 [hep-th]}
  \BibitemShut {NoStop}%
\bibitem [{\citenamefont {Branchina}\ \emph {et~al.}(2003)\citenamefont
  {Branchina}, \citenamefont {Meissner},\ and\ \citenamefont
  {Veneziano}}]{Branchina:2003ek}%
  \BibitemOpen
  \bibfield  {author} {\bibinfo {author} {\bibfnamefont {V.}~\bibnamefont
  {Branchina}}, \bibinfo {author} {\bibfnamefont {K.~A.}\ \bibnamefont
  {Meissner}}, \ and\ \bibinfo {author} {\bibfnamefont {G.}~\bibnamefont
  {Veneziano}},\ }\href {\doibase 10.1016/j.physletb.2003.09.020} {\bibfield
  {journal} {\bibinfo  {journal} {Phys. Lett.}\ }\textbf {\bibinfo {volume}
  {B574}},\ \bibinfo {pages} {319} (\bibinfo {year} {2003})},\ \Eprint
  {http://arxiv.org/abs/hep-th/0309234} {arXiv:hep-th/0309234 [hep-th]}
  \BibitemShut {NoStop}%
\bibitem [{\citenamefont {Reuter}(1996)}]{Reuter:1996ub}%
  \BibitemOpen
  \bibfield  {author} {\bibinfo {author} {\bibfnamefont {M.}~\bibnamefont
  {Reuter}},\ }in\ \href@noop {} {\emph {\bibinfo {booktitle} {{5th Hellenic
  School and Workshops on Elementary Particle Physics (CORFU 1995) Corfu,
  Greece, September 3-24, 1995}}}}\ (\bibinfo {year} {1996})\ \Eprint
  {http://arxiv.org/abs/hep-th/9602012} {arXiv:hep-th/9602012 [hep-th]}
  \BibitemShut {NoStop}%
\bibitem [{\citenamefont {Percacci}(2007)}]{Percacci:2007sz}%
  \BibitemOpen
  \bibfield  {author} {\bibinfo {author} {\bibfnamefont {R.}~\bibnamefont
  {Percacci}},\ }\href@noop {} {\  (\bibinfo {year} {2007})},\ \Eprint
  {http://arxiv.org/abs/0709.3851} {arXiv:0709.3851 [hep-th]} \BibitemShut
  {NoStop}%
\bibitem [{\citenamefont {Nagy}(2014)}]{Nagy:2012ef}%
  \BibitemOpen
  \bibfield  {author} {\bibinfo {author} {\bibfnamefont {S.}~\bibnamefont
  {Nagy}},\ }\href {\doibase 10.1016/j.aop.2014.07.027} {\bibfield  {journal}
  {\bibinfo  {journal} {Annals Phys.}\ }\textbf {\bibinfo {volume} {350}},\
  \bibinfo {pages} {310} (\bibinfo {year} {2014})},\ \Eprint
  {http://arxiv.org/abs/1211.4151} {arXiv:1211.4151 [hep-th]} \BibitemShut
  {NoStop}%
\bibitem [{\citenamefont {Demmel}\ and\ \citenamefont
  {Nink}(2015)}]{Demmel:2015zfa}%
  \BibitemOpen
  \bibfield  {author} {\bibinfo {author} {\bibfnamefont {M.}~\bibnamefont
  {Demmel}}\ and\ \bibinfo {author} {\bibfnamefont {A.}~\bibnamefont {Nink}},\
  }\href {\doibase 10.1103/PhysRevD.92.104013} {\bibfield  {journal} {\bibinfo
  {journal} {Phys. Rev.}\ }\textbf {\bibinfo {volume} {D92}},\ \bibinfo {pages}
  {104013} (\bibinfo {year} {2015})},\ \Eprint
  {http://arxiv.org/abs/1506.03809} {arXiv:1506.03809 [gr-qc]} \BibitemShut
  {NoStop}%
\bibitem [{\citenamefont {Ohta}\ \emph
  {et~al.}(2016{\natexlab{b}})\citenamefont {Ohta}, \citenamefont {Percacci},\
  and\ \citenamefont {Pereira}}]{Ohta:2016npm}%
  \BibitemOpen
  \bibfield  {author} {\bibinfo {author} {\bibfnamefont {N.}~\bibnamefont
  {Ohta}}, \bibinfo {author} {\bibfnamefont {R.}~\bibnamefont {Percacci}}, \
  and\ \bibinfo {author} {\bibfnamefont {A.~D.}\ \bibnamefont {Pereira}},\
  }\href {\doibase 10.1007/JHEP06(2016)115} {\bibfield  {journal} {\bibinfo
  {journal} {JHEP}\ }\textbf {\bibinfo {volume} {06}},\ \bibinfo {pages} {115}
  (\bibinfo {year} {2016}{\natexlab{b}})},\ \Eprint
  {http://arxiv.org/abs/1605.00454} {arXiv:1605.00454 [hep-th]} \BibitemShut
  {NoStop}%
\bibitem [{\citenamefont {Litim}(2001)}]{Litim:2001up}%
  \BibitemOpen
  \bibfield  {author} {\bibinfo {author} {\bibfnamefont {D.~F.}\ \bibnamefont
  {Litim}},\ }\href {\doibase 10.1103/PhysRevD.64.105007} {\bibfield  {journal}
  {\bibinfo  {journal} {Phys.Rev.}\ }\textbf {\bibinfo {volume} {D64}},\
  \bibinfo {pages} {105007} (\bibinfo {year} {2001})},\ \Eprint
  {http://arxiv.org/abs/hep-th/0103195} {arXiv:hep-th/0103195 [hep-th]}
  \BibitemShut {NoStop}%
\bibitem [{\citenamefont {Benedetti}(2012)}]{Benedetti:2011ct}%
  \BibitemOpen
  \bibfield  {author} {\bibinfo {author} {\bibfnamefont {D.}~\bibnamefont
  {Benedetti}},\ }\href {\doibase 10.1088/1367-2630/14/1/015005} {\bibfield
  {journal} {\bibinfo  {journal} {New J. Phys.}\ }\textbf {\bibinfo {volume}
  {14}},\ \bibinfo {pages} {015005} (\bibinfo {year} {2012})},\ \Eprint
  {http://arxiv.org/abs/1107.3110} {arXiv:1107.3110 [hep-th]} \BibitemShut
  {NoStop}%
\bibitem [{xAc()}]{xActwebpage}%
  \BibitemOpen
  \href@noop {} {\enquote {\bibinfo {title} {{xAct: Efficient tensor computer
  algebra for Mathematica}},}\ }\bibinfo {howpublished}
  {\url{http://xact.es/index.html}},\ \bibinfo {note} {accessed:
  2015-07-30}\BibitemShut {NoStop}%
\bibitem [{\citenamefont {Brizuela}\ \emph {et~al.}(2009)\citenamefont
  {Brizuela}, \citenamefont {Martin-Garcia},\ and\ \citenamefont
  {Mena~Marugan}}]{Brizuela:2008ra}%
  \BibitemOpen
  \bibfield  {author} {\bibinfo {author} {\bibfnamefont {D.}~\bibnamefont
  {Brizuela}}, \bibinfo {author} {\bibfnamefont {J.~M.}\ \bibnamefont
  {Martin-Garcia}}, \ and\ \bibinfo {author} {\bibfnamefont {G.~A.}\
  \bibnamefont {Mena~Marugan}},\ }\href {\doibase 10.1007/s10714-009-0773-2}
  {\bibfield  {journal} {\bibinfo  {journal} {Gen. Rel. Grav.}\ }\textbf
  {\bibinfo {volume} {41}},\ \bibinfo {pages} {2415} (\bibinfo {year}
  {2009})},\ \Eprint {http://arxiv.org/abs/0807.0824} {arXiv:0807.0824 [gr-qc]}
  \BibitemShut {NoStop}%
\bibitem [{\citenamefont
  {{Mart{\'{\i}}n-Garc{\'{\i}}a}}(2008)}]{2008CoPhC.179..597M}%
  \BibitemOpen
  \bibfield  {author} {\bibinfo {author} {\bibfnamefont {J.~M.}\ \bibnamefont
  {{Mart{\'{\i}}n-Garc{\'{\i}}a}}},\ }\href {\doibase
  10.1016/j.cpc.2008.05.009} {\bibfield  {journal} {\bibinfo  {journal}
  {Computer Physics Communications}\ }\textbf {\bibinfo {volume} {179}},\
  \bibinfo {pages} {597} (\bibinfo {year} {2008})},\ \Eprint
  {http://arxiv.org/abs/0803.0862} {arXiv:0803.0862 [cs.SC]} \BibitemShut
  {NoStop}%
\bibitem [{\citenamefont {{Mart{\'{\i}}n-Garc{\'{\i}}a}}\ \emph
  {et~al.}(2007)\citenamefont {{Mart{\'{\i}}n-Garc{\'{\i}}a}}, \citenamefont
  {{Portugal}},\ and\ \citenamefont {{Manssur}}}]{2007CoPhC.177..640M}%
  \BibitemOpen
  \bibfield  {author} {\bibinfo {author} {\bibfnamefont {J.~M.}\ \bibnamefont
  {{Mart{\'{\i}}n-Garc{\'{\i}}a}}}, \bibinfo {author} {\bibfnamefont
  {R.}~\bibnamefont {{Portugal}}}, \ and\ \bibinfo {author} {\bibfnamefont
  {L.~R.~U.}\ \bibnamefont {{Manssur}}},\ }\href {\doibase
  10.1016/j.cpc.2007.05.015} {\bibfield  {journal} {\bibinfo  {journal}
  {Computer Physics Communications}\ }\textbf {\bibinfo {volume} {177}},\
  \bibinfo {pages} {640} (\bibinfo {year} {2007})},\ \Eprint
  {http://arxiv.org/abs/0704.1756} {arXiv:0704.1756 [cs.SC]} \BibitemShut
  {NoStop}%
\bibitem [{\citenamefont {{Mart{\'{\i}}n-Garc{\'{\i}}a}}\ \emph
  {et~al.}(2008)\citenamefont {{Mart{\'{\i}}n-Garc{\'{\i}}a}}, \citenamefont
  {{Yllanes}},\ and\ \citenamefont {{Portugal}}}]{2008CoPhC.179..586M}%
  \BibitemOpen
  \bibfield  {author} {\bibinfo {author} {\bibfnamefont {J.~M.}\ \bibnamefont
  {{Mart{\'{\i}}n-Garc{\'{\i}}a}}}, \bibinfo {author} {\bibfnamefont
  {D.}~\bibnamefont {{Yllanes}}}, \ and\ \bibinfo {author} {\bibfnamefont
  {R.}~\bibnamefont {{Portugal}}},\ }\href {\doibase 10.1016/j.cpc.2008.04.018}
  {\bibfield  {journal} {\bibinfo  {journal} {Computer Physics Communications}\
  }\textbf {\bibinfo {volume} {179}},\ \bibinfo {pages} {586} (\bibinfo {year}
  {2008})},\ \Eprint {http://arxiv.org/abs/0802.1274} {arXiv:0802.1274 [cs.SC]}
  \BibitemShut {NoStop}%
\bibitem [{\citenamefont {{Nutma}}(2014)}]{2014CoPhC.185.1719N}%
  \BibitemOpen
  \bibfield  {author} {\bibinfo {author} {\bibfnamefont {T.}~\bibnamefont
  {{Nutma}}},\ }\href {\doibase 10.1016/j.cpc.2014.02.006} {\bibfield
  {journal} {\bibinfo  {journal} {Computer Physics Communications}\ }\textbf
  {\bibinfo {volume} {185}},\ \bibinfo {pages} {1719} (\bibinfo {year}
  {2014})},\ \Eprint {http://arxiv.org/abs/1308.3493} {arXiv:1308.3493 [cs.SC]}
  \BibitemShut {NoStop}%
\bibitem [{\citenamefont {Barvinsky}\ and\ \citenamefont
  {Vilkovisky}(1985)}]{Barvinsky:1985an}%
  \BibitemOpen
  \bibfield  {author} {\bibinfo {author} {\bibfnamefont {A.~O.}\ \bibnamefont
  {Barvinsky}}\ and\ \bibinfo {author} {\bibfnamefont {G.~A.}\ \bibnamefont
  {Vilkovisky}},\ }\href {\doibase 10.1016/0370-1573(85)90148-6} {\bibfield
  {journal} {\bibinfo  {journal} {Phys. Rept.}\ }\textbf {\bibinfo {volume}
  {119}},\ \bibinfo {pages} {1} (\bibinfo {year} {1985})}\BibitemShut {NoStop}%
\bibitem [{\citenamefont {Decanini}\ and\ \citenamefont
  {Folacci}(2006)}]{Decanini:2005gt}%
  \BibitemOpen
  \bibfield  {author} {\bibinfo {author} {\bibfnamefont {Y.}~\bibnamefont
  {Decanini}}\ and\ \bibinfo {author} {\bibfnamefont {A.}~\bibnamefont
  {Folacci}},\ }\href {\doibase 10.1103/PhysRevD.73.044027} {\bibfield
  {journal} {\bibinfo  {journal} {Phys. Rev.}\ }\textbf {\bibinfo {volume}
  {D73}},\ \bibinfo {pages} {044027} (\bibinfo {year} {2006})},\ \Eprint
  {http://arxiv.org/abs/gr-qc/0511115} {arXiv:gr-qc/0511115 [gr-qc]}
  \BibitemShut {NoStop}%
\bibitem [{\citenamefont {Anselmi}\ and\ \citenamefont
  {Benini}(2007)}]{Anselmi:2007eq}%
  \BibitemOpen
  \bibfield  {author} {\bibinfo {author} {\bibfnamefont {D.}~\bibnamefont
  {Anselmi}}\ and\ \bibinfo {author} {\bibfnamefont {A.}~\bibnamefont
  {Benini}},\ }\href {\doibase 10.1088/1126-6708/2007/10/099} {\bibfield
  {journal} {\bibinfo  {journal} {JHEP}\ }\textbf {\bibinfo {volume} {10}},\
  \bibinfo {pages} {099} (\bibinfo {year} {2007})},\ \Eprint
  {http://arxiv.org/abs/0704.2840} {arXiv:0704.2840 [hep-th]} \BibitemShut
  {NoStop}%
\bibitem [{\citenamefont {Groh}\ \emph
  {et~al.}(2011{\natexlab{b}})\citenamefont {Groh}, \citenamefont
  {Saueressig},\ and\ \citenamefont {Zanusso}}]{Groh:2011dw}%
  \BibitemOpen
  \bibfield  {author} {\bibinfo {author} {\bibfnamefont {K.}~\bibnamefont
  {Groh}}, \bibinfo {author} {\bibfnamefont {F.}~\bibnamefont {Saueressig}}, \
  and\ \bibinfo {author} {\bibfnamefont {O.}~\bibnamefont {Zanusso}},\
  }\href@noop {} {\  (\bibinfo {year} {2011}{\natexlab{b}})},\ \Eprint
  {http://arxiv.org/abs/1112.4856} {arXiv:1112.4856 [math-ph]} \BibitemShut
  {NoStop}%
\end{thebibliography}%

\end{document}